\documentclass[sigconf]{acmart}

\AtBeginDocument{%
  \providecommand\BibTeX{{%
    \normalfont B\kern-0.5em{\scshape i\kern-0.25em b}\kern-0.8em\TeX}}}

\usepackage[linesnumbered, ruled, vlined]{algorithm2e}
 
\SetAlFnt{\small}
\SetAlCapFnt{\small}
\SetAlCapNameFnt{\small}
\SetAlCapHSkip{0pt}
\IncMargin{-\parindent}
\SetKwProg{Fn}{Function}{}{end}
\SetKwFunction{CACHEMODEL}{cacheModel}
\SetKwFunction{PATHTORMO}{path2RMO}
\SetKwFunction{RMOTOPATH}{RMO2path}
\SetKwFunction{INCACHE}{inCache}
\SetKwFunction{ADDBLOCK}{addBlock2Cache}

\acmDOI{}
\acmISBN{}
\acmConference[Arxiv Preprint]{Arxiv Preprint}{July 2023}{Chattanooga,
  TN, USA}%
\begin{document}

\title{The Impact of Space-Filling Curves  on   Data Movement in 
Parallel Systems}

\author{David Walker}
\authornote{Both authors contributed equally to this research.}
\email{david-walker02@utc.edu}
\orcid{0000-0002-1360-6330}
\author{Anthony Skjellum}
\authornotemark[1]
\email{tony-skjellum@utc.edu}
\affiliation{%
  \streetaddress{Department of Computer Science and Engineering}
  \institution{University of Tennessee at Chattanooga}
  \city{Chattanooga}
  \state{Tennessee}
  \country{USA}
  \postcode{37403}
}

\renewcommand{\shortauthors}{Walker and Skjellum}

\begin{abstract}
  Modern computer systems are  characterised by deep memory hierarchies, composed of main memory, multiple layers of cache, and other specialised types of memory. In parallel and distributed systems, additional memory layers are added to this hierarchy. Achieving good performance for computational science applications, in terms of execution time, depends on the efficient use of this diverse and hierarchical memory. This paper revisits the use of space-filling curves to specify the ordering in memory of data structures used in representative scientific applications executing on parallel machines containing clusters of multicore CPUs with attached GPUs. This work examines the hypothesis that space-filling curves, such as Hilbert and Morton ordering, can improve data locality and hence result in more efficient data movement than row or column-based orderings. First, performance results are presented that show for what application parameterisations and machine characteristics this is the case, and are interpreted in terms of how an application interacts with the computer hardware and low-level software. This research particularly focuses on the use 
  of stencil-based applications that form the basis of many scientific computations. Second, how space-filling curves impact data sharing in nearest-neighbour and stencil-based codes is considered. 
\end{abstract}


\maketitle

\section{Introduction}
In the majority of modern high performance computers used for scientific computing, data movement rather than floating-point computation is the main factor in determining overall execution time. Moreover, typically such systems are highly heterogeneous, being composed of multiple types of processor and levels of hierarchical memory that interact in complex ways, often through architecture-specific mechanisms. Thus, achieving performance portability in parallel algorithms and applications across existing and anticipated high performance computing (HPC) resources is a challenge. Exploiting data locality is essential in the efficient use of complex hierarchical memory. However, such memories generally are not under direct programmatic control, and so good data locality must be coerced by judicious choice of data access patterns, which in turn depend on how data are stored in memory and order of computation. This style of programming is exemplified by the use of Level 3 BLAS operations in numerical linear algebra computations~\cite{BLAS3}, which make use of block matrix operations to achieve efficient use of cache. Another related approach to performance portability is the use of autotuning in which data layout and other application parameters are optimised dynamically at runtime, usually in a preprocessing step~\cite{CLINTWHALEY20013}. 

This paper revisits the use of space-filling
curves to specify the ordering in memory of data structures used
in representative scientific applications executing on parallel machines.
This research extends our previous work~\cite{IJHPCA_Walker,AlKharusiWalker2019} to examine the hypothesis that space-filling curves, such as
Hilbert and Morton\footnote{Strictly speaking Morton order does not correspond to a curve as it lacks the necessary continuity requirement, but this distinction is not relevant to this work} ordering, can improve data locality and hence
result in more efficient data movement than row or column-based
orderings. Space-filling curves recursively generate  data orderings, with each level in the recursion corresponding to contiguous blocks of data that fit into particular levels of the memory hierarchy, hence reducing the cost of data movement. This topic is investigated for the following situations:
\begin{enumerate}
    \item Updating data at  locations in a 3D array using data within a 3D stencil, as is commonly used in scientific computations. In this case, we investigate whether updating data by following a particular path through the 3D data volume results in better computational efficiency if such a path results in more efficient use of the memory hierarchy.
    \item Packing/unpacking data into/from a communication buffer when transferring data associated (via MPI or otherwise) with the surfaces of a 3D data array between processes in a parallel application. For a row-major or column-major ordering, different faces can be packed and unpacked with differing efficiencies because data in different surfaces are accessed with different strides. The impact of strided access in such cases is expected to be less for data ordered by a space-filling curve.
\end{enumerate} 
In all cases, row-major, Hilbert, and Morton data orderings are compared. 

The rest of this paper is structured as follows. Section~\ref{section:related} presents previous work on the use of space-filling curves, and describes the Hilbert and Morton orderings. 
In Section~\ref{section:methodology},  the methodology used in this investigation is described. Experimental performance results are presented in Section~\ref{section:results} and are analysed and interpreted in Section~\ref{section:analysis}. Finally, in Section~\ref{section:conclusions} conclusions and directions for future work are presented.

\section{Related Work}
\label{section:related}

Morton ordering~\cite{Morton1966} has been used to optimize database access, in image processing algorithms, and in dense linear algebra computations~ \cite{Chatterjee1999,LortonWise,TBK}. The use of Morton ordering in a number of matrix multiplication algorithms, including Strassen’s algorithm, has also been investigated by Valsalam and Skjellum for an earlier generation of processors~\cite{Valsalam2002}. More recently Morton ordering has been used in tensor computations~\cite{PAWLOWSKI201934}. DeFord and Kayanaraman~\cite{Deford2013} have investigated the use of space-filling curves in mapping data to processes in parallel applications. Wide-ranging uses of space-filling curves in scientific computing have been presented by Bader~\cite{bader}.  The use of the Hilbert  and Morton ordering in data layout has been investigated for molecular dynamics applications~\cite{MellorCrummey2001,AlKharusiWalker2019}.

\subsection{Morton Ordering}
\label{subsection:morton}
\begin{figure}[ht]
  \centering
  \includegraphics[width=\linewidth]{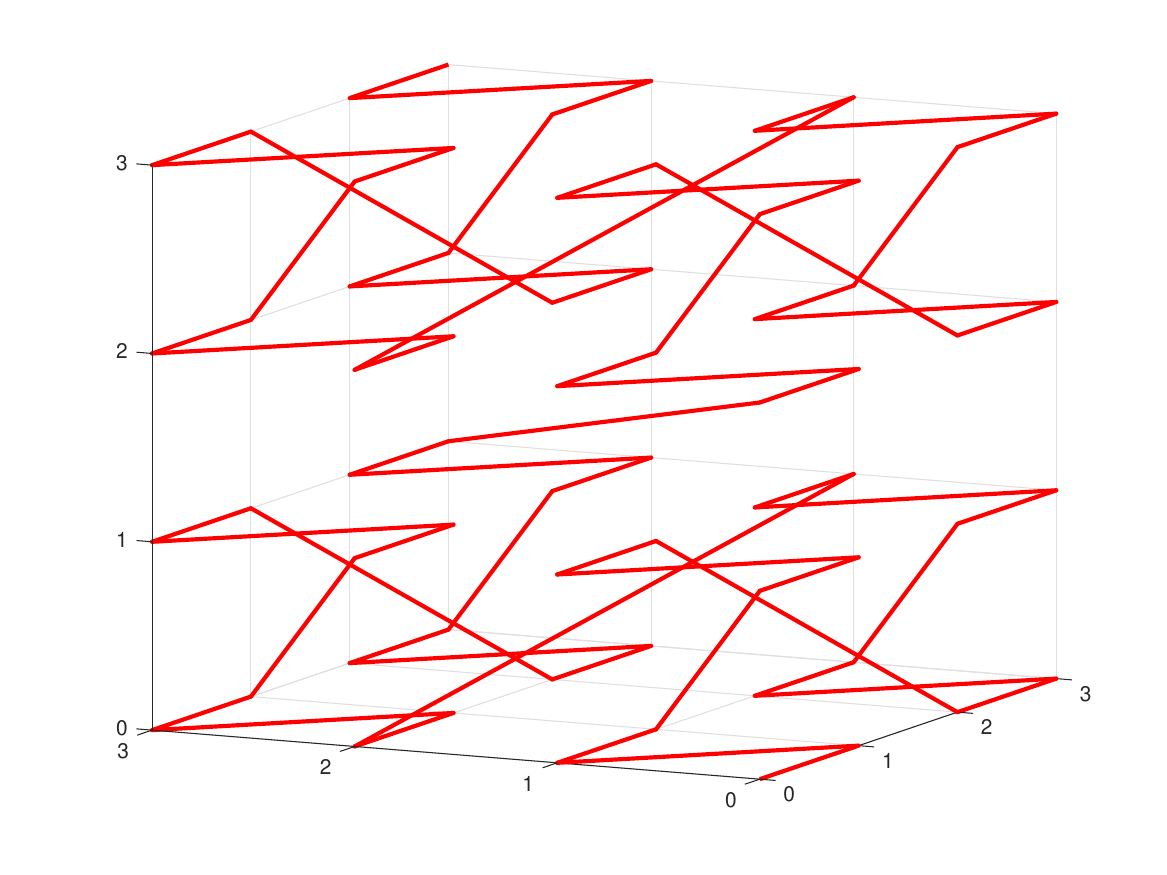}
  \caption{Three-dimensional Morton ordering for a $4\times 4\times 4$ array. The ordering starts at $(0,0,0)$ and ends at $(3,3,3)$.}
  \label{fig:morton3D}
\end{figure}

Morton ordering, $\mathcal{O}_M$, takes a 3D array stored in row-major order and re-orders it as a $2\times 2\times 2$ block array in which the items of each block is stored in row-major order. This process can
then be applied recursively to each of the eight blocks, and after $r$ levels of recursion, the array will be reordered as $8r$ sub-arrays, each in row-major order. 
A similar approach can be applied to arrays in column-major order.
Morton ordering can be applied to arbitrary arrays; however, for the rest of this article, attention will focus on Morton ordering of $M\times M\times M$ arrays, where $M = 2^m$. Applying Morton ordering to such an array to level $r$ results in sub-arrays of size $2^{m-r}\times 2^{m-r}\times 2^{m-r}$. Level $r = 0$ corresponds to the original array, and so $0\le r< m$. If $r = m -1$, the Morton blocks are of minimum size, namely $2\times 2\times 2$. Figure~\ref{fig:morton3D} illustrates this case for a $4\times 4 \times 4$ array of data.

\begin{figure}[ht]
  \centering
  \includegraphics[width=\linewidth]{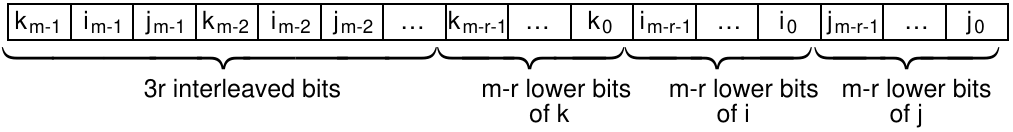}
  \caption{The bits of the Morton index $\ell_r$ for a 3D mapping at level $r\ge 1$.}
  \label{fig:mortonbits}
\end{figure}

Applying Morton ordering to a depth $r$ can be expressed as a manipulation of the bitwise representation of the array indices, $(k,i,j)$, to give the Morton index, $\ell_r$. Here, $j$ is the column index, $i$ the row index, and $k$ the slab index. The upper $r$ bits of $k$, $i$, and $j$ are interleaved  to form the upper $3r$ bits of $\ell_r$. The lower $m-r$ bits of $k$ form the next least significant bits of $\ell_r$, followed by the lower $m-r$ bits of $i$. Finally, the lower $m-r$ bits of $j$ form the least significant bits of $\ell_r$. This is shown in Fig.~\ref{fig:mortonbits}. The sub-arrays defined by Morton ordering can be numbered consecutively from 0 according to the order in which they are visited. The interleaved upper $r$ bits of $k$. $i$, and $j$ give the number of the sub-array containing $(k,i,j)$, while the lower $m-r$ bits of $k$, $i$, and $j$ give the location within the 3D sub-array. 
Given a Morton ordering at level $r-1$, the ordering at level $r$ is obtained by applying the following two bitwise rotations to the Morton index $\ell_{r-1}$:
\begin{enumerate}
    \item Cyclically rotate bits $2(m-r)$ to $3(t-r)$ to the right.
    \item Cyclically rotate bits $(m-r)$ to $3(t-r)$ to the right.
\end{enumerate}
These operations generate $\ell_r$ from $\ell_{r-1}$.

Raman and Wise~\cite{RamanWise2007} have shown how Morton ordering of 2D arrays can be performed using dilated integers to interleave the bits that make up the $3r$ most significant bits of $\ell_r$. The technique used here is a simple extension of their approach.

\subsection{Hilbert Ordering}
\begin{figure}[ht]
  \centering
  \includegraphics[width=\linewidth]{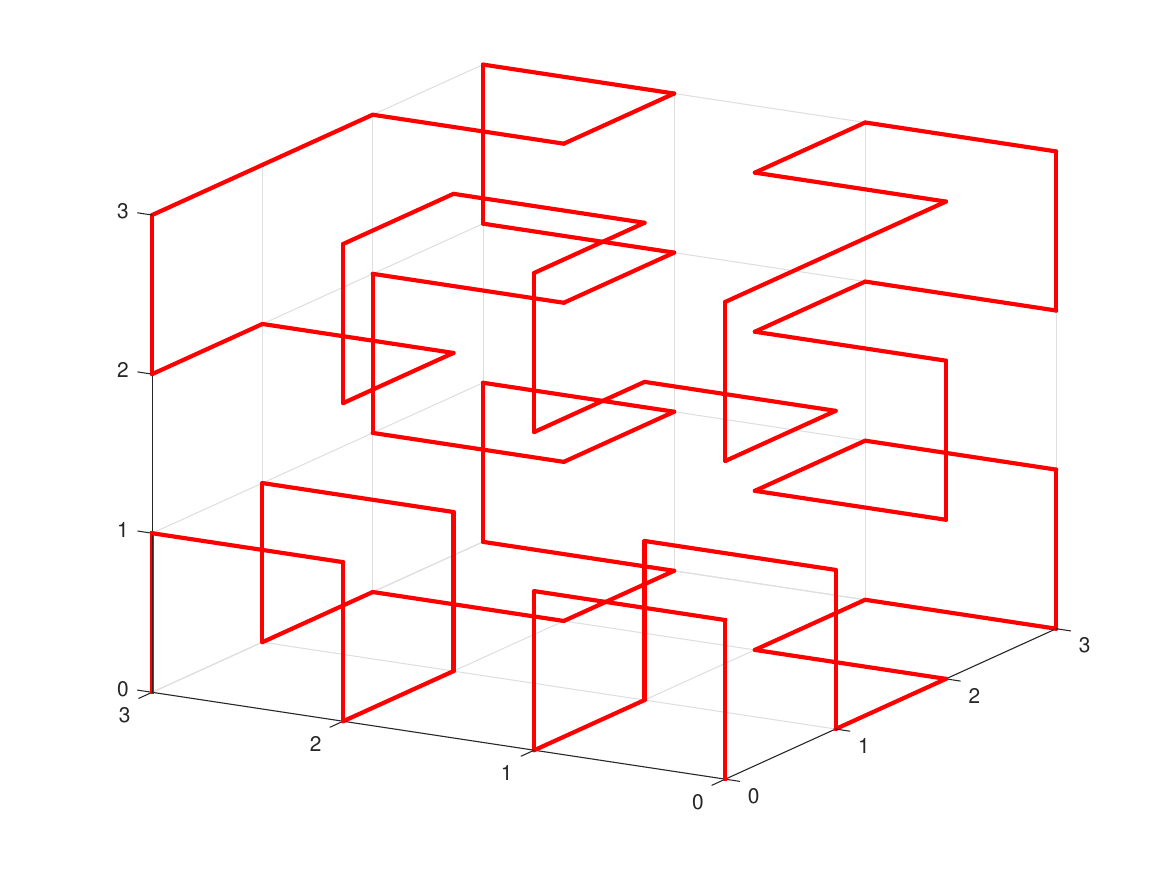}
  \caption{Three-dimensional Hilbert ordering for a $4\times 4\times 4$ array. The ordering starts at $(0,0,0)$ and ends at $(3,3,3)$.}
  \label{fig:hilbert3D}
\end{figure}

A Hilbert ordering, $\mathcal{O}_H$, follows the path of a space-filling Hilbert curve through an $M\times M\times M$ array. The Hilbert ordering requires that $M=2^m$ for some $m\ge 2$. Figure~\ref{fig:hilbert3D} shows the Hilbert ordering for a $4\times 4 \times 4$ array of data.

The mapping between array location, $(k,i,j)$, in the 3D array and the Hilbert index can be expressed in terms of a Lindenmayer system~\cite{Lindenmayer1968,Prusinkiewicz1996}, as described in~\cite{AlKharusiWalker2019}.

\subsection{Hybrid Orderings}
Hybrid orderings are obtained by splitting the 3D array of size $M\times M\times M$ into sub-arrays of equal size $T\times T\times T$, and applying one ordering within the sub-arrays and another ordering between them. A hybrid ordering that uses row-major ordering within the sub-arrays and Morton ordering between them has been discussed above in Section~\ref{subsection:morton}.
Another example, would be to apply a Hilbert ordering within the sub-arrays and
a row-major ordering between them
(provided $T=2^t$ for some $t\ge 2$). In general, if the sub-arrays are cubes of size $2^t$ the lower $3t$ bits give the position in the sub-array and the upper $3(m-t)$ bits give the index of the Morton or Hilbert ordering between the sub-arrays.


\section{Methodology}
\label{section:methodology}
A three-dimensional block of data is considered, containing an array of $M\times M\times M$ data items of equal size. 

\subsection{Stencil-Based Computation}
Suppose that processing the data at position $(k,i,j)$ in the array accesses once data within a stencil that is a cube of size $2g+1$ in each direction, centered on $(k,i,j)$. An indication of how efficiently data are accessed within an hierarchical memory can be gained by investigating the offsets in memory required to process each data item. This can be done by evaluating the number of memory accesses for a given memory offset, $x$, and a given data ordering, $\mathcal{O}$, when processing data in an array of size $(M-2g)\times(M-2g)\times(M-2g)$, which will be denoted by $h_{\mathcal{O}}(x)$. Denote by $n_{\mathcal{O}}(x;k,i,j)$ the number of memory offsets of size $x$ for a stencil centred on $(k,i,j)$. Then,
\[
h_\mathcal{O}(x)=\sum_{k,i,j,}n_{\mathcal{O}}(x;k,i,j)
\]
where the summation is over all stencils that fit entirely within the $M\times M\times M$ array. Thus, $g\le k,i,j <M-g$.

\begin{figure}[ht]
  \centering
  \includegraphics[width=\linewidth]{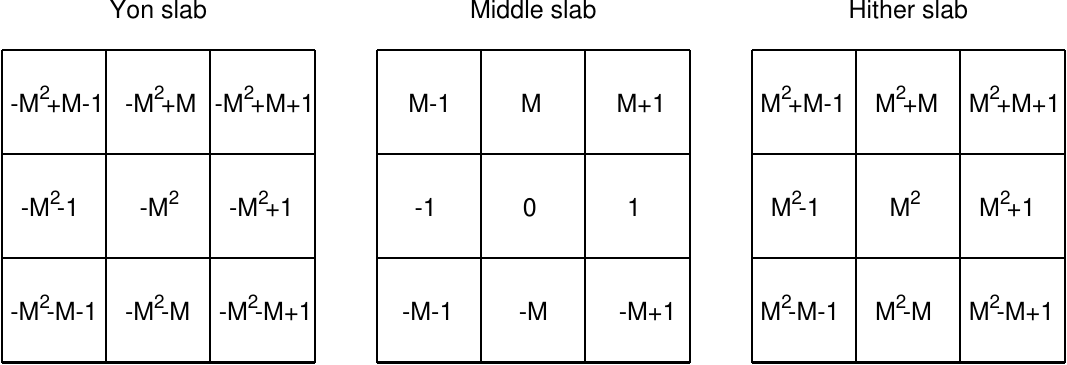}
  \caption{Memory offsets for a row-major ordering for a stencil with $g=1$. For a column-major ordering the slabs should be rotated 90 degrees clockwise.}
  \label{fig:offsets}
\end{figure}

\begin{figure}[ht]
  \centering
  \includegraphics[width=\linewidth]{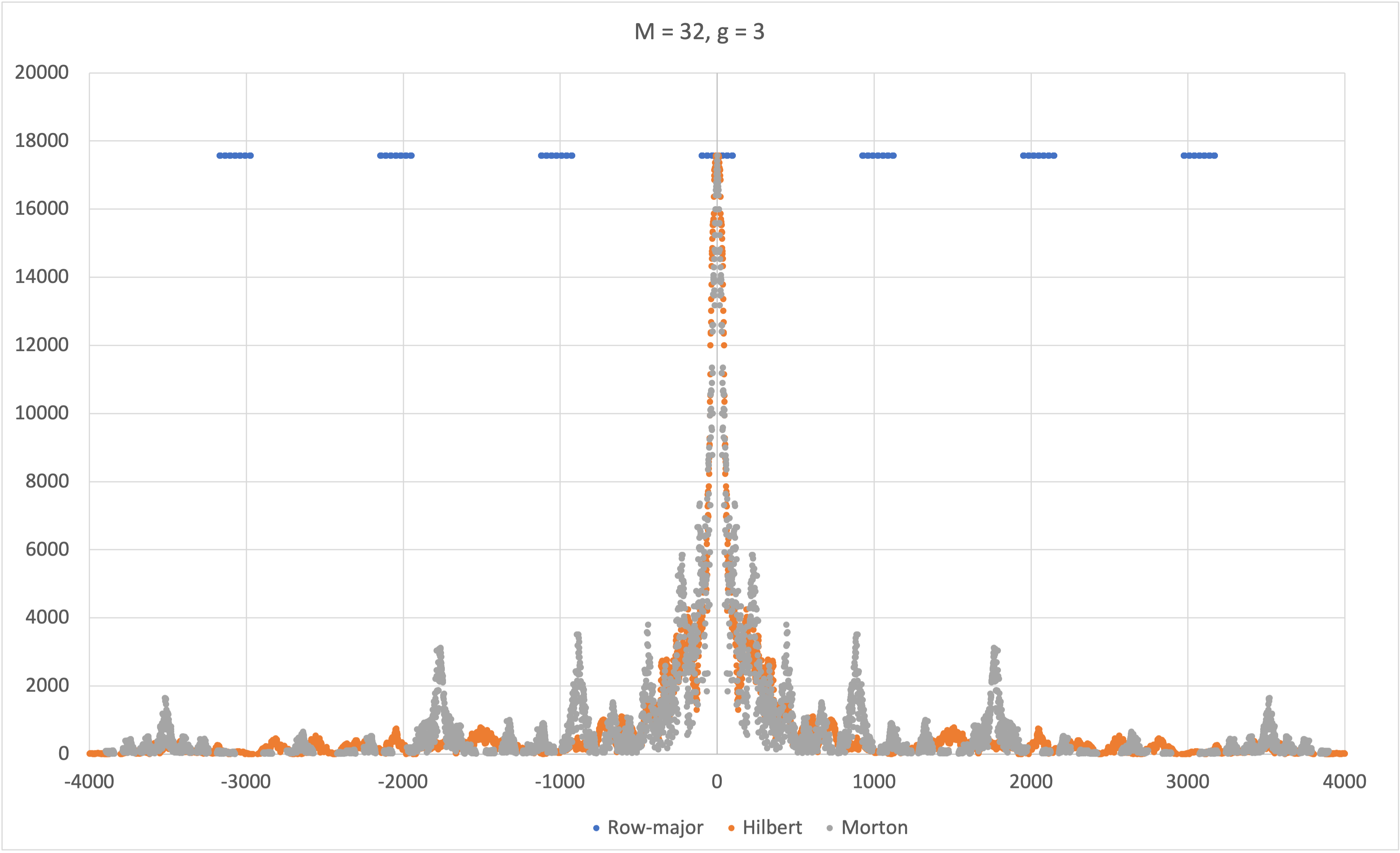}
  \caption{Accumulated memory offsets for a stencil with $g=1$ and an array with $M=32$.}
  \label{fig:h1}
\end{figure}

\begin{figure}[ht]
  \centering
  \includegraphics[width=\linewidth]{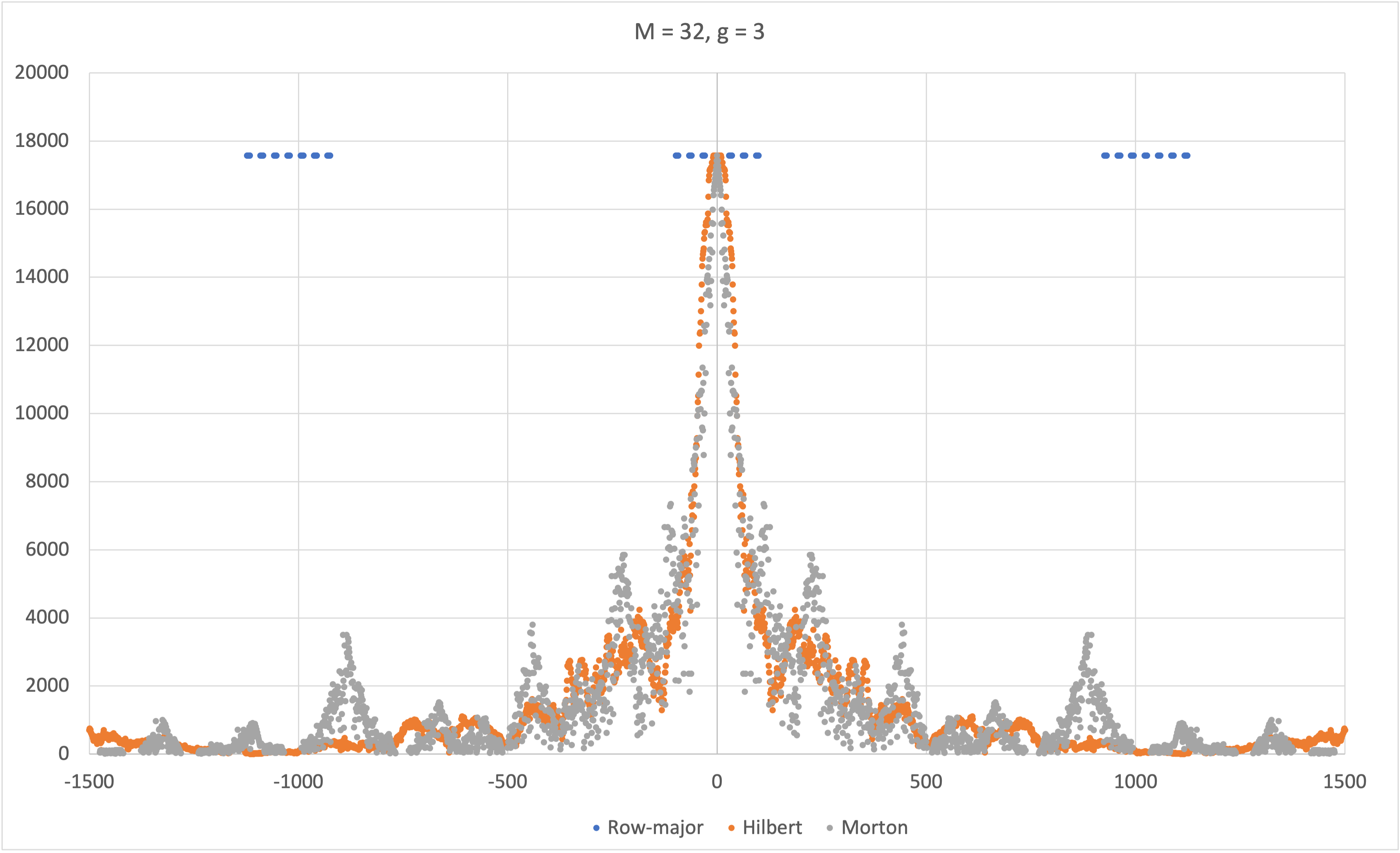}
  \caption{Accumulated memory offsets for a stencil with $g=3$ and an array with $M=32$.}
  \label{fig:h2}
\end{figure}

\begin{figure}[ht]
  \centering
  \includegraphics[width=\linewidth]{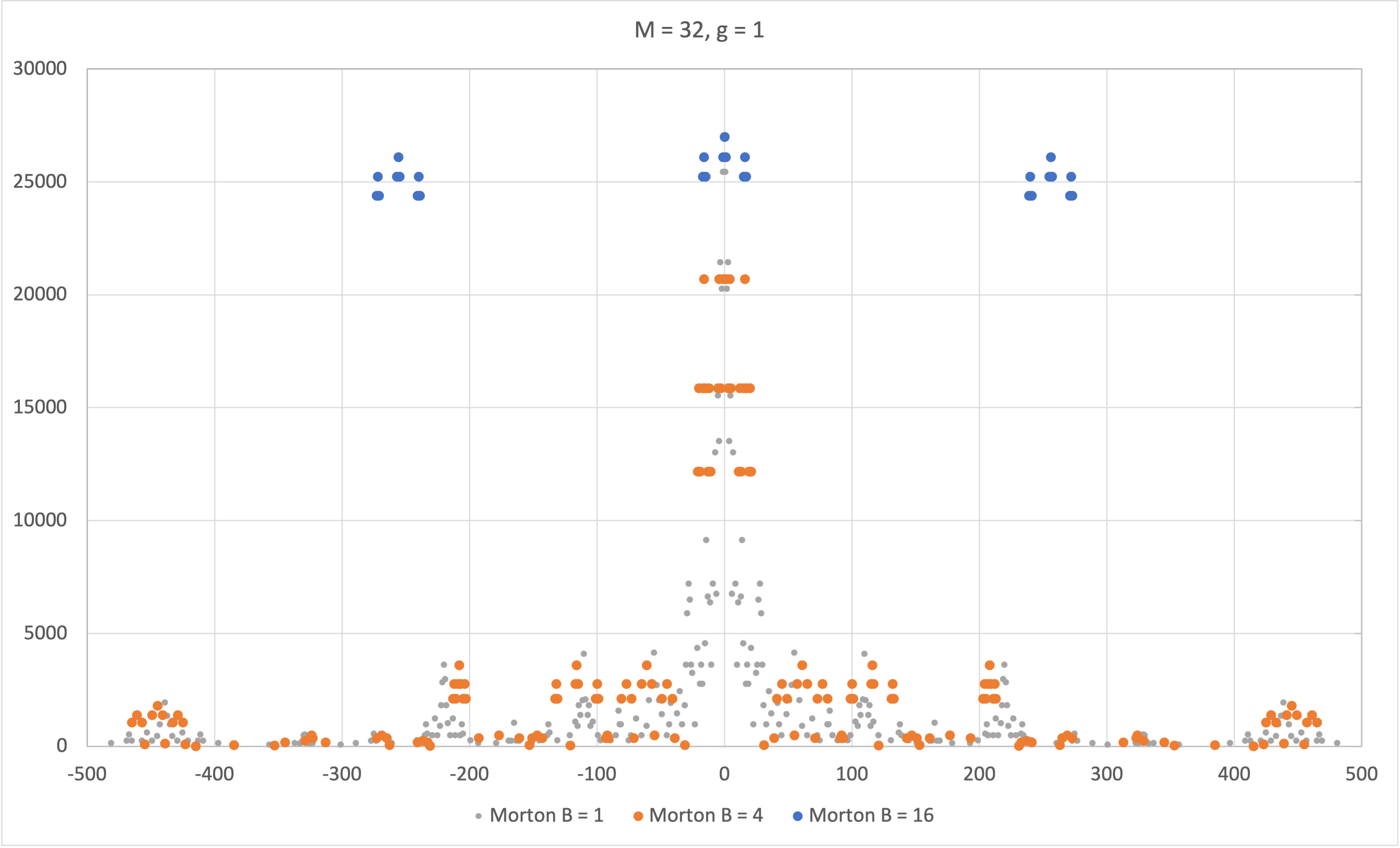}
  \caption{Accumulated memory offsets for a stencil with $g=1$ and an array with $M=32$ for a Morton ordering with block sizes 1, 4, and 16.}
  \label{fig:MortonBlocked}
\end{figure}

For a row or column major ordering $n_{\mathcal{O}}(x;k,i,j)$ is independent of $(k,i,j)$. For a stencil with $g=1$, the corresponding memory offsets at each stencil location for a row-major ordering are shown in Fig.~\ref{fig:offsets}. In general, for a row or column major ordering, there are $(2g+1)^3$ memory offsets, $x$, for which $h_{\mathcal{O}}(x)$ is $(M-2g)^3$, and $h_{\mathcal{O}}(x)$ is zero for all other memory offsets. However, for Hilbert and Morton orderings, $n_{\mathcal{O}}(x;k,i,j)$ depends on  the  stencil location, $(k,i,j)$. For all orderings, the histogram $h_{\mathcal{O}}(x)$ can be found computationally. For example, Figs.~\ref{fig:h1} and \ref{fig:h2} show the histograms for $g=1$ and $g=3$, with $M=32$ in both cases. Figure~\ref{fig:MortonBlocked} shows how varying the block size in a Morton ordering affects the accumulated memory offsets for $M=32$ and $g=1$. Clearly, there is a greater scatter in the memory access patterns for the Hilbert and Morton orderings, and this extends beyond the x-axis in Figs.~\ref{fig:h1}-\ref{fig:MortonBlocked}.

To further investigate how the memory access patterns generated by different orderings might affect performance, a simple cache model has been developed. In this model memory (at some level in the memory hierarchy) is divided into cache lines, each of which can hold $b$ data items. This memory  can contain up to $c$ cache lines. Clearly, this model only captures the dynamics of two levels of memory. A cache miss occurs whenever a data item is not held in memory, and in this case a cache line containing the missing data is transferred (from another level in the memory hierarchy or from storage) into the memory. If there is insufficient space in the memory to receive the cache line, then the least recently used cache line is ejected from memory to make space. Each location in the data cube (except for those in the border region of width $g$) is accessed in the specified order. For each data item the corresponding stencil locations are accessed and the number of cache misses is recorded.

The parameters of the cache model for a given ordering, are the stencil size, $g$, the number of data items in each dimension, $M$, the number of data items, $b$, in a cache line, and the number of cache lines, $c$, that fit into the memory. Pseudocode expressing the high level structure of the cache model is given in Alg.~\ref{alg:cachemodel}.

\begin{algorithm}[ht]
\SetAlgoNoLine
\DontPrintSemicolon
\Fn{\CACHEMODEL{ordering,M,g,b,c}} {
\KwIn{$\mathit{ordering}$, integers $M$ and $g$ defining size of the array and the stencil, and integers $b$ and $c$ defining the size of a cache line and the number of cache lines in memory.}
\KwOut{The number of cache misses $\mathit{nmisses}$.}
$\mathit{nmisses}$ = $0$\;
\ForEach{{\rm(location,} {ipath},  {\rm in} ordering{\rm )}}{
$\mathit{irmo}$ = \PATHTORMO($\mathit{ipath}$)\;
\If{{\rm (}irmo {\rm not in border zone)}}{
\ForEach{{\rm (stencil offset,} soff{\rm )}}{
$\mathit{jrmo}$ = $\mathit{ibin}+\mathit{soff}$\;
$\mathit{jpath}$ = \RMOTOPATH($\mathit{jrmo}$)\;
\If{(\rm{!}\INCACHE($\mathit{jpath}$))}{
    $\mathit{nmisses}$++\;
    \ADDBLOCK($\mathit{jpath}$)\;
}
}
}
}
return $\mathit{nmisses}$\;
}
\caption{cacheModel: high level view of the cache model. The functions \texttt{path2RMO} and \texttt{RMO2path} convert between a location in the ordering and the row-major index.}
\label{alg:cachemodel}
\end{algorithm}

\subsection{Accessing Data in Surfaces of the Data Cube}
Denote by $(k,i,j)$ a location at slab $k$, row $i$, and column $j$ of the data cube. We further denote by $(0:g-1,:,:)$ the $g\times M\times M$ front surface of the data cube of depth $g$ spanned by rows and columns. $(M-g:M-1,:,:)$ is the corresponding back surface of the data cube. The front and back surfaces spanned by columns and slabs are denoted by $(:,0:g-1,:)$ and $(:,M-g:M-1,:)$, respectively. Finally, $(:,:,0:g-1)$ and $(:,:,M-g:M-1)$ are the front and back surfaces spanned by slabs and rows.

Let $p(k,i,j)$ be the path index at location $(k,i,j)$, that is, the position in the row-major, Morton, or Hilbert ordering. Let $q(r)$ be the inverse mapping; that is, $q(r)$ is the position in a row-major ordering of position $r$ in the chosen ordering, from which $(k,i,j)$ can be deduced.

Consider one of the six surfaces, $S$, and let $p_t$ ($t=0,1,\ldots,gM^2-1)$ be the positions in the ordering for points in the surface as we follow the path through the data cube given by the ordering. The proximity in memory of locations in the surface can be assessed using a variant of the cache model in Alg.~\ref{alg:cachemodel}. All that is required is for the conditional statement in line 5 to be negated.



\section{Experimental Performance Results}
\label{section:results}
Having gained some insights into the impact of data ordering on the use of hierarchical memory, we next examine how the performance of a 3D stencil-based C++ code named {\it gol3d} is affected by different row-major, Hilbert, and Morton orderings. The {\it gold3d} code extends the Game of Life simulation~\cite{gol1970} by allowing the size of the stencil (the cubical region that determines how a data location is updated) to be specified at execution time.

The performance experiments were carried out on two processors:
\begin{enumerate}
    \item 
A 2.0GHz AMD Epyc 7662 processor with 64 cores, a 32k L1 data cache, a 32k L1 instruction cache, a 512k L2 cache, and a 16384k L3 cache.  
\item
A 2.9GHz Intel Xeon Gold 6226R processor with 16 cores and a 22 MB smart cache (a proprietary technology for which further details are not publicly available).
\end{enumerate}
In all cases the code was compiled with gcc 4.8.5 and the -O2 optimization flag set (it was found that using the -O3 optimization flag resulted in no significant improvement in performance). Each value reported is the average of 10 measured values.

For each processor, two sets of timing experiments were performed for row-major, Morton, and Hilbert orderings. In the first set the performance of {\it gol3d} was investigated by timing how long it took to perform 10 updates of the 3D grid. The order in which  data points were updated is given by the path of the ordering through the data cube.

In the second set of experiments, the impact of data order on communication between processes in a parallel version of {\it gol3d} was investigated. The key determinant for this is the time to pack (and unpack) values in the surface of the data cube into a contiguous communication buffers. There are six surfaces and for a stencil width $g$ each of the corresponding data buffers contain $gn^2$ grid values. This type of "halo" communication is common in spatially decomposed parallel applications, and in iterative and time-stepping problems may be performed many thousands of times in a simulation. To avoid unnecessary repetition of index calculations, an initial traversal of the data path (determined by the ordering) is made and lists of path indices in each surface region are made at a memory cost of $6gn^2$ integers. These are then used to efficiently pack the buffers.

\subsection{Eypc Results}
\begin{figure}[ht]
  \centering
  \includegraphics[width=\linewidth]{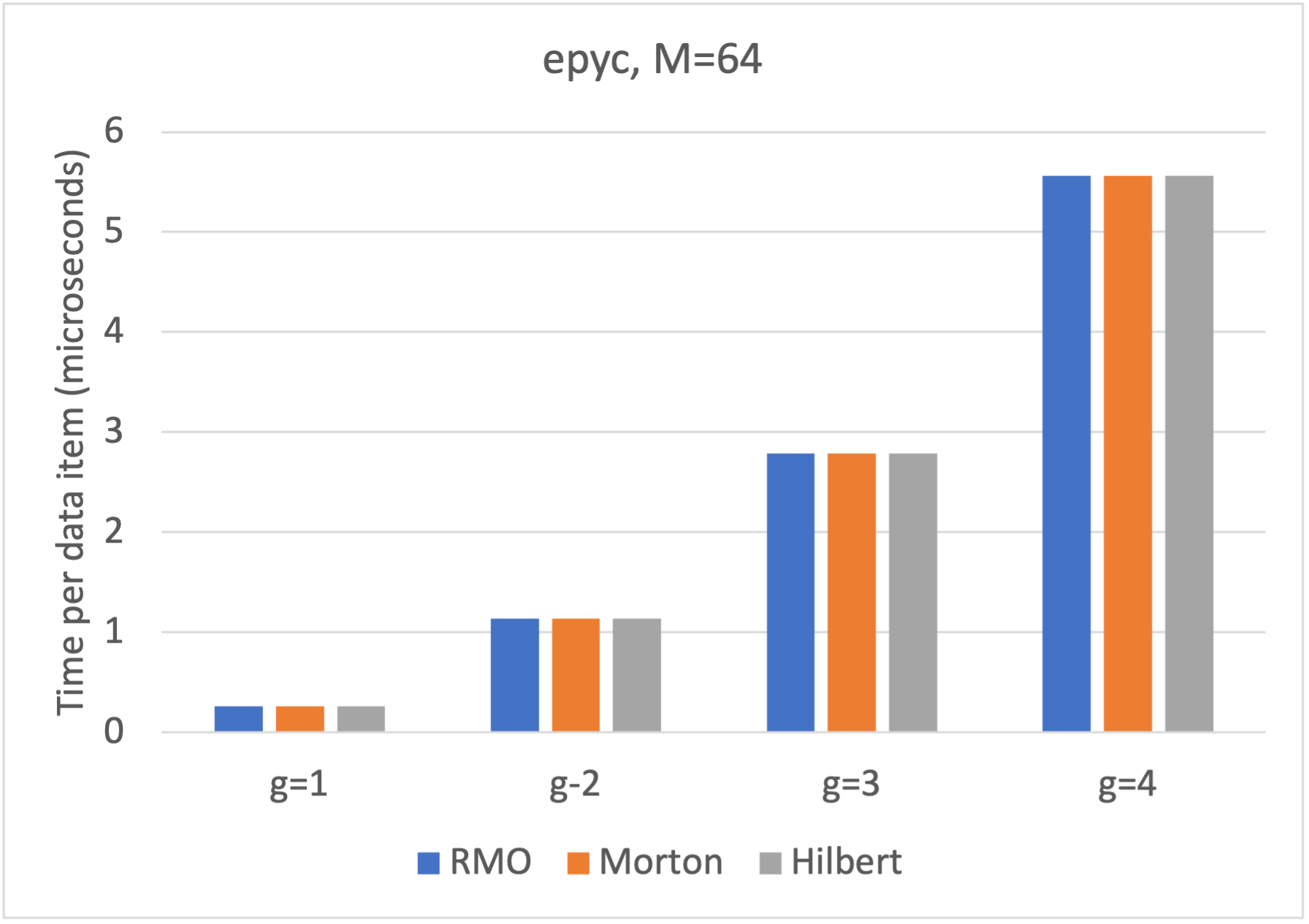}
  \caption{Time per data item for 10 iterations for a $64\times 64\times 64$ grid for different orderings and stencil sizes, $g=1, 2, 3$, and $4$.}
  \label{fig:epyc_gol3d_n64}
\end{figure}

\begin{figure}[ht]
  \centering
  \includegraphics[width=\linewidth]{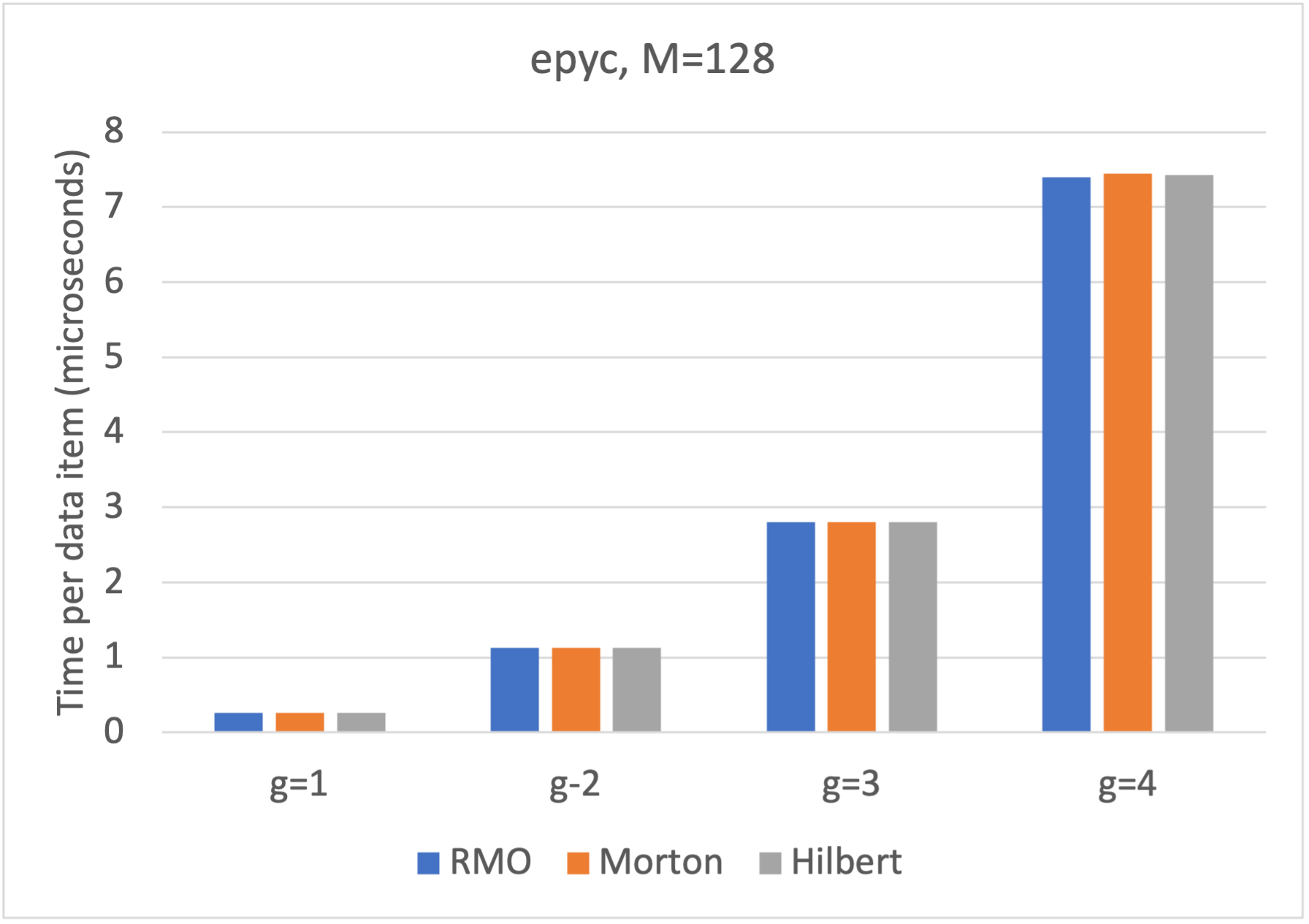}
  \caption{Time per data item for 10 iterations for a $128\times 128\times 128$ grid for different orderings and stencil sizes, $g=1, 2, 3$, and $4$.}
  \label{fig:epyc_gol3d_n128}
\end{figure}

\begin{figure}[ht]
  \centering
  \includegraphics[width=\linewidth]{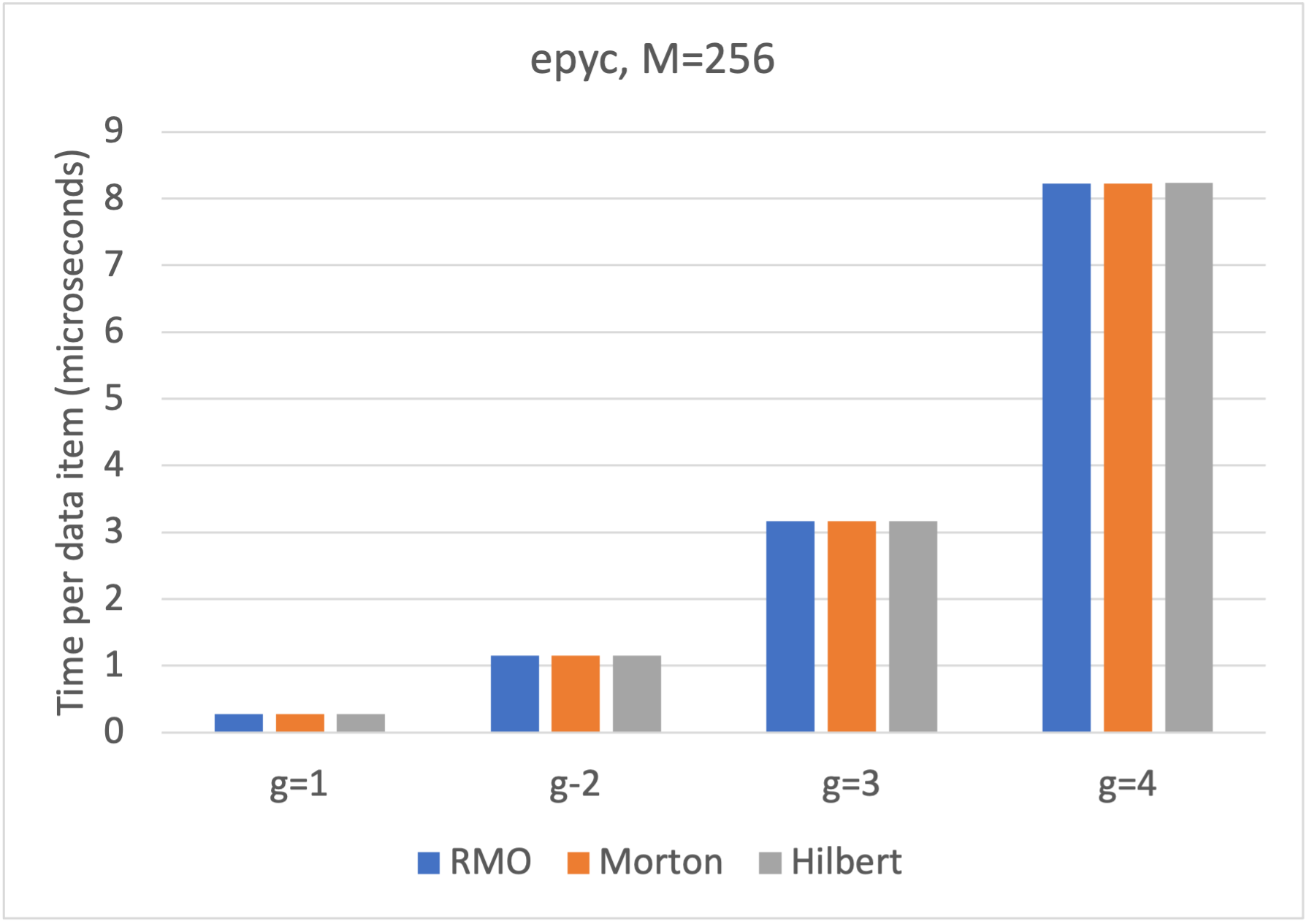}
  \caption{Time per data item for 10 iterations for a $256\times 256\times 256$ grid for different orderings and stencil sizes, $g=1, 2, 3$, and $4$.}
  \label{fig:epyc_gol3d_n256}
\end{figure}

 Figures~\ref{fig:epyc_gol3d_n64}-\ref{fig:epyc_gol3d_n256} present results for running {\it gol3d} on the Epyc processor. It is clear from these figures that the ordering has no impact on the time to update a grid value. For a given stencil size the time per grid value update  depends weakly on the problem size, $M$,  for the $g=4$, and is independent of $M$ for smaller values of $g$. The times shown scale approximately with stencil size, $(2g+1)^3$. One standard deviation error bars are not plotted in the figures because the variation in runtime between different executions is so small that they would not be visible.

\begin{figure*}[ht]
  \includegraphics[width=0.4\linewidth]{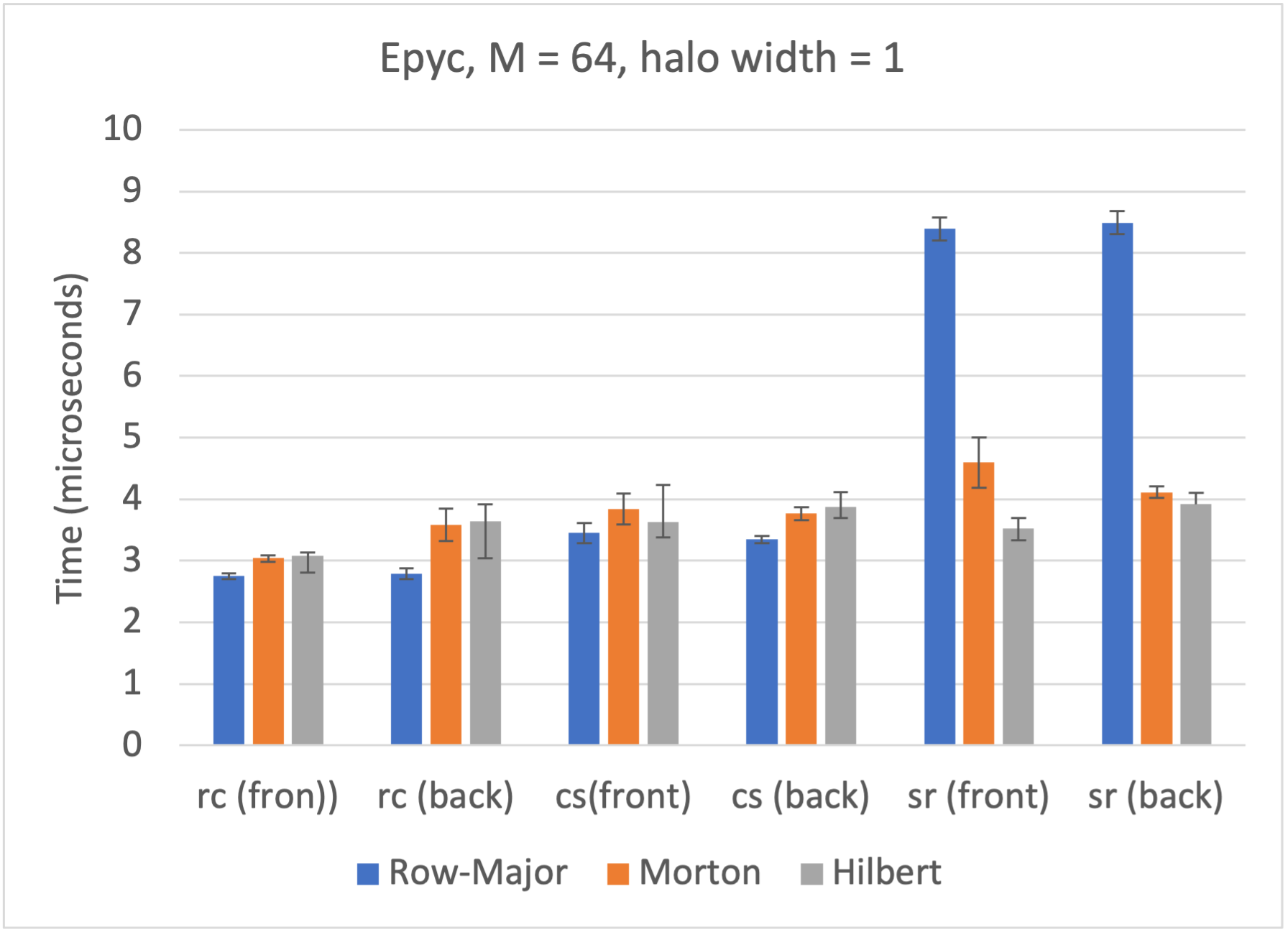}
  \qquad
  \includegraphics[width=0.4\linewidth]{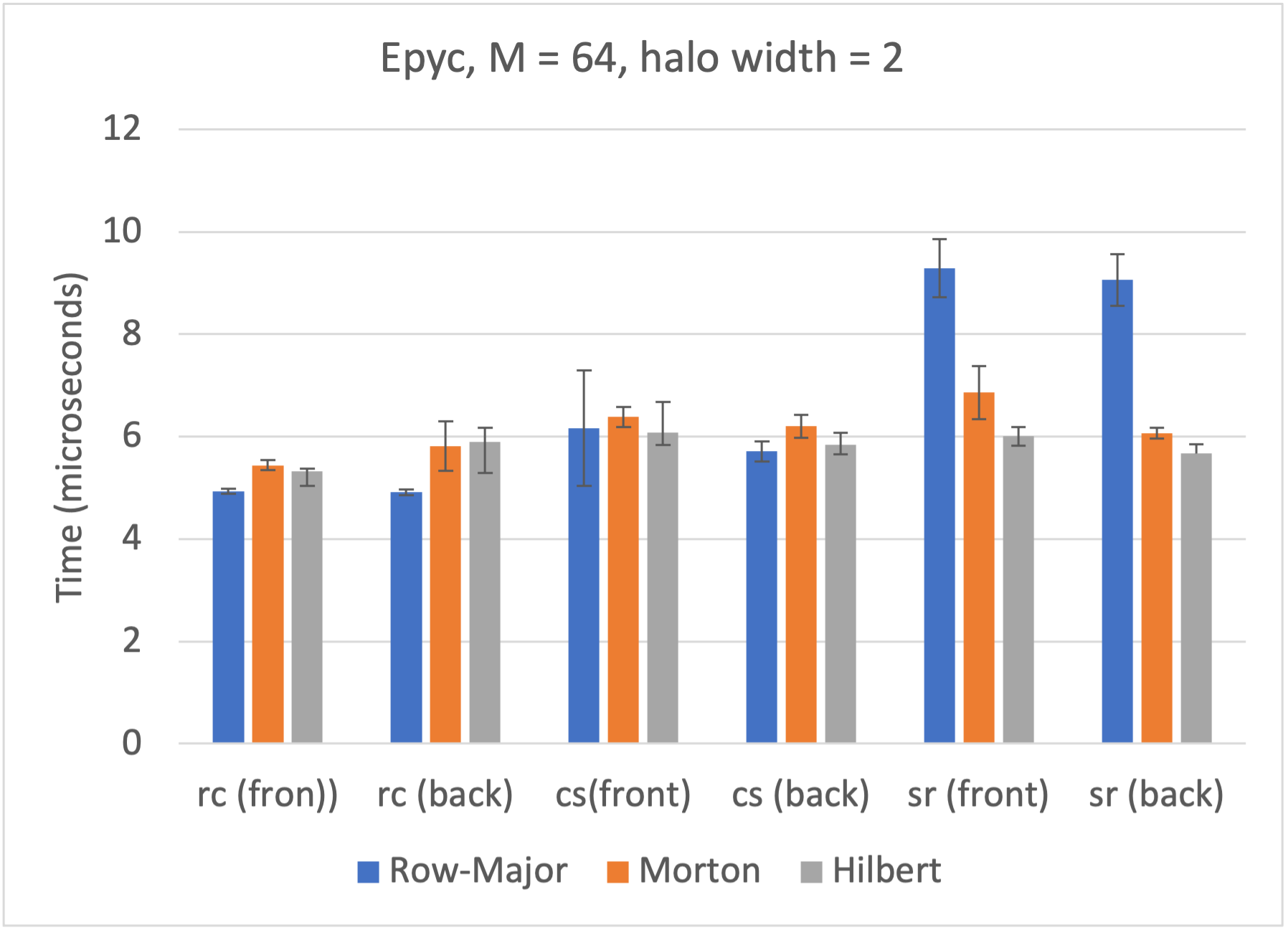}  \\
    \includegraphics[width=0.4\linewidth]{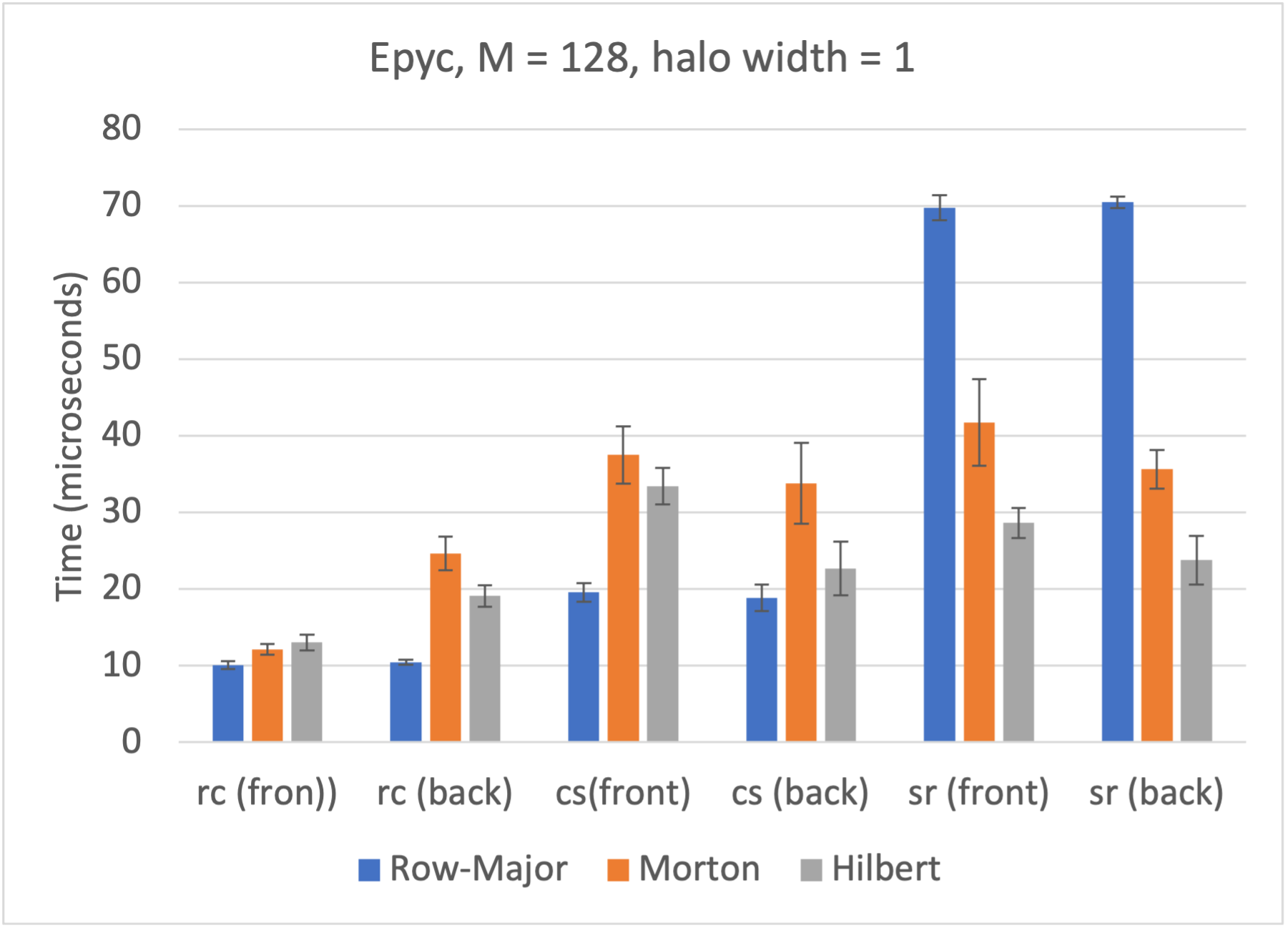}
  \qquad
  \includegraphics[width=0.4\linewidth]{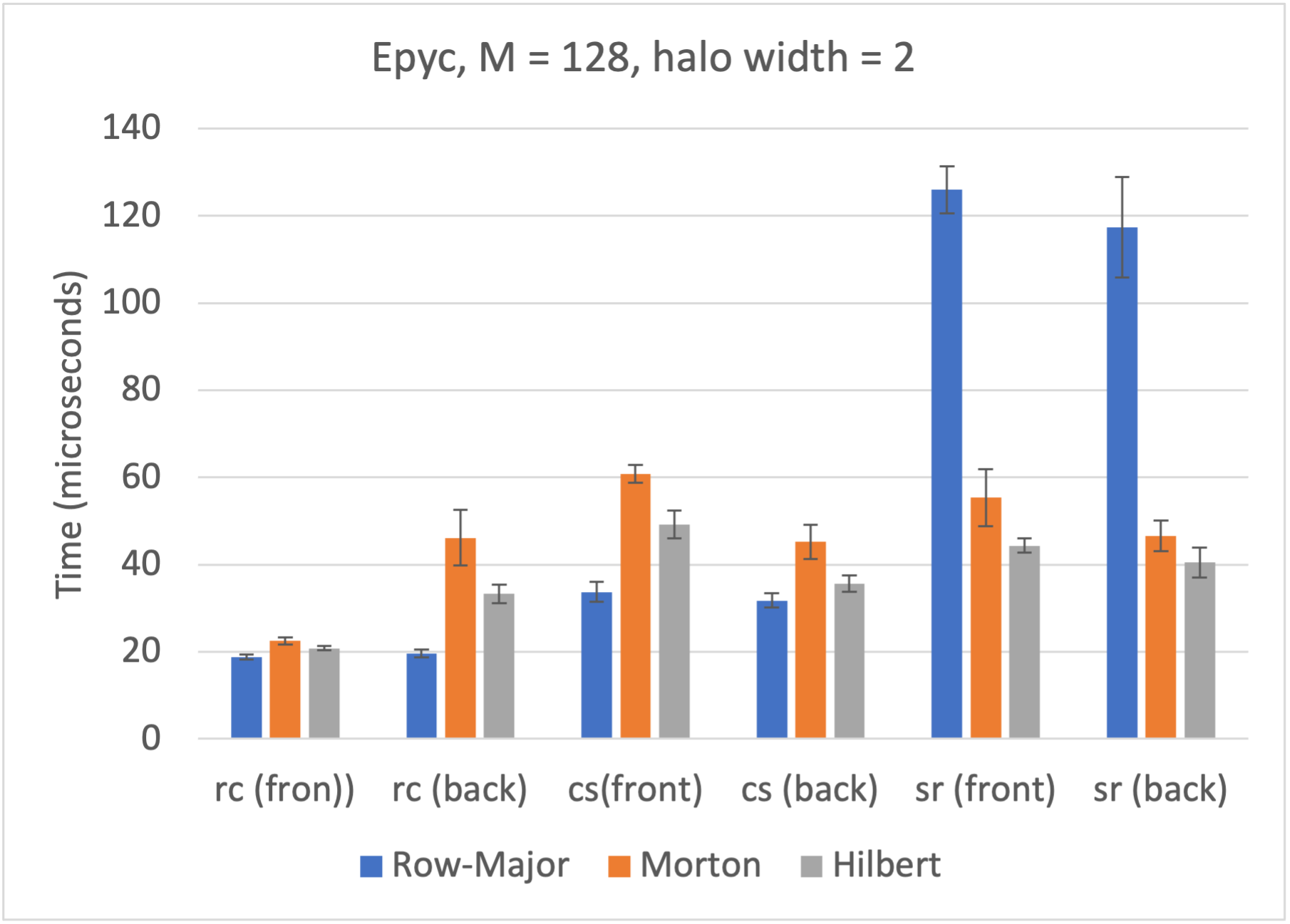} \\
      \includegraphics[width=0.4\linewidth]{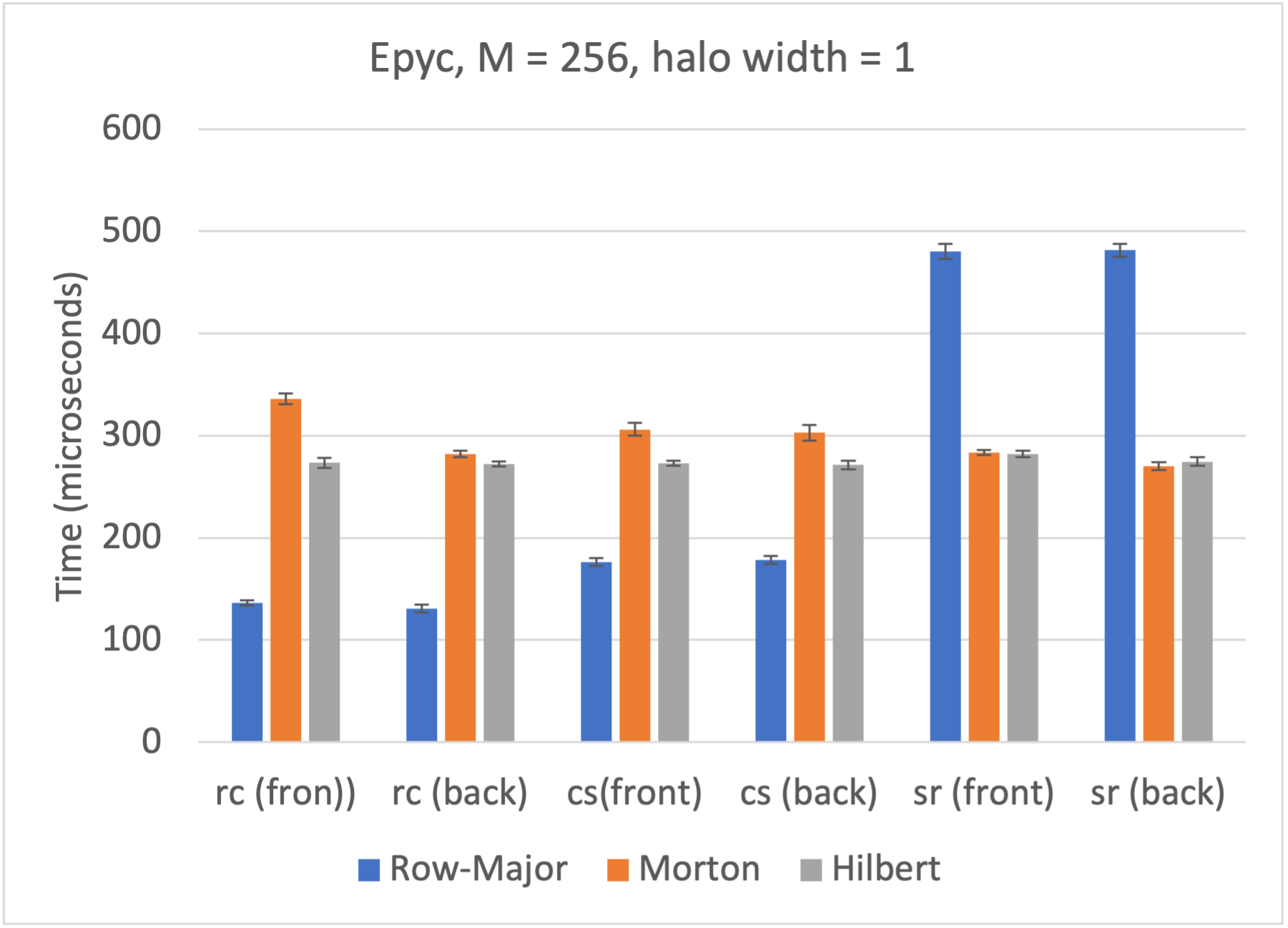}
  \qquad
  \includegraphics[width=0.4\linewidth]{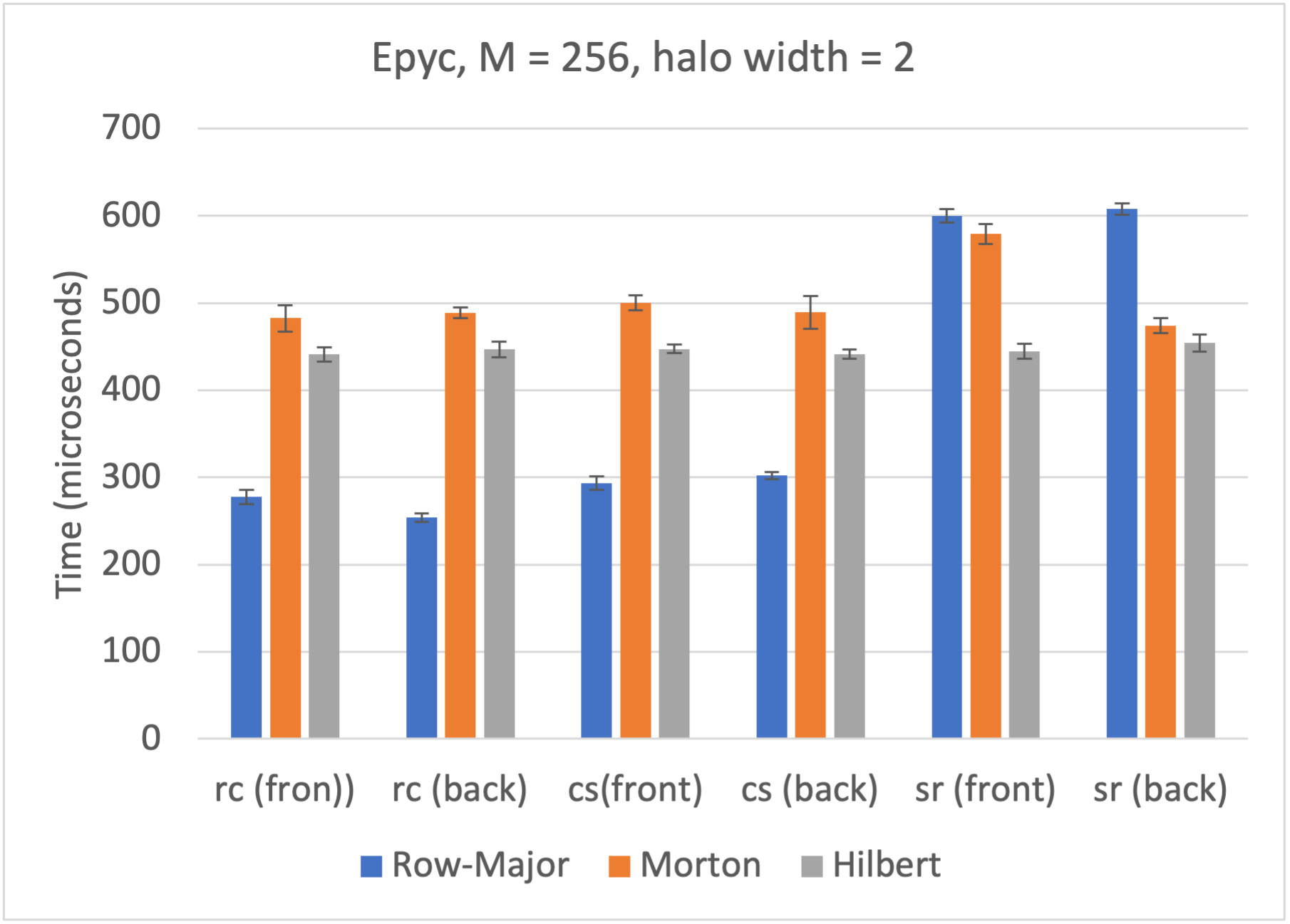}  \caption{Epyc results: time to buffer surfaces, for halo widths 1 and 2 and problem sizes 64, 128, and 256.}
  \label{fig:epyc_buffer}

\end{figure*}

Representative results of the time to pack a communication buffer with data from the surfaces of the data cube are shown in Fig.~\ref{fig:epyc_buffer}

\subsection{Results for Xeon}
\begin{figure}[ht]
  \centering
  \includegraphics[width=\linewidth]{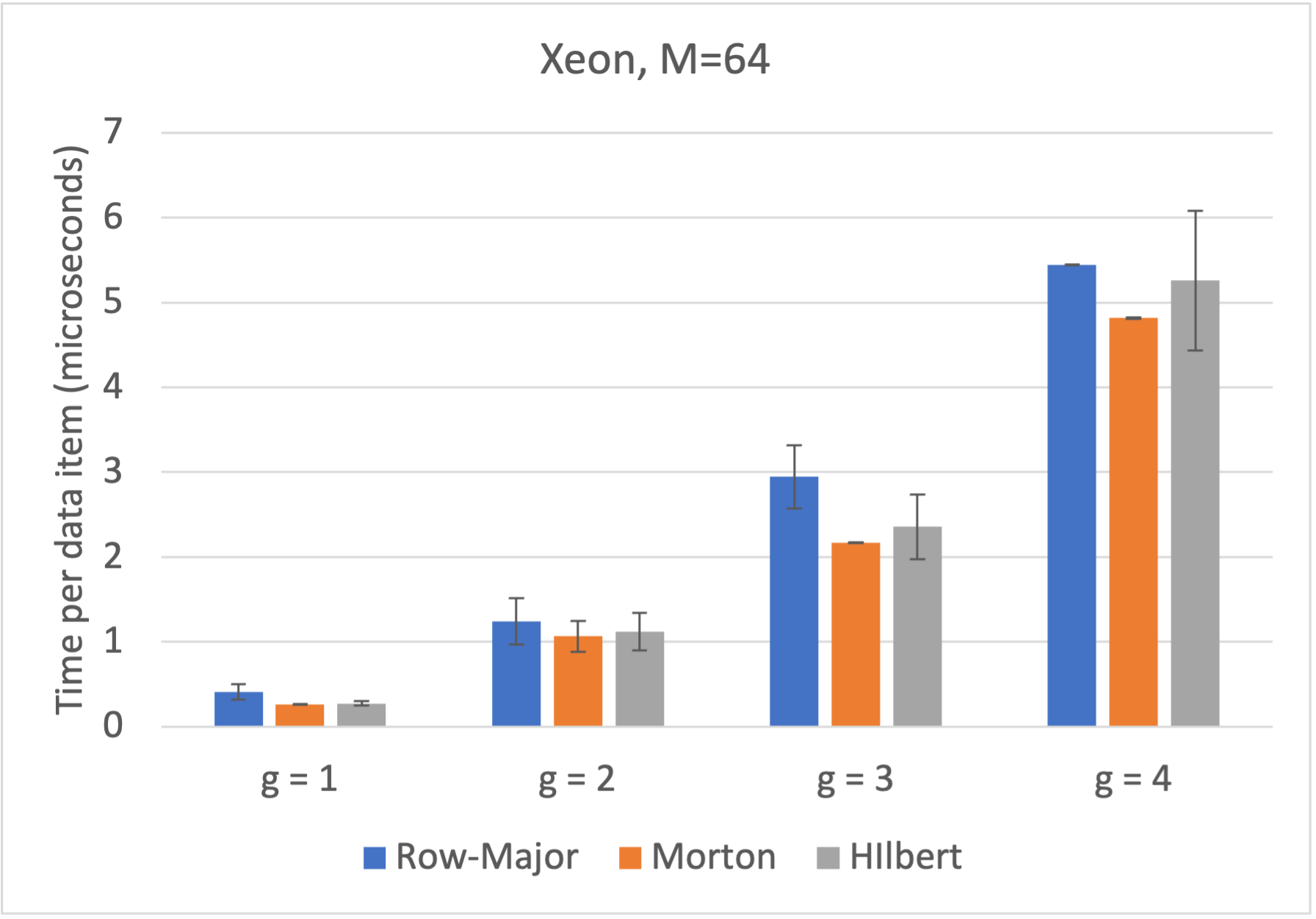}
  \caption{Time per data item for 10 iterations for a $64\times 64\times 64$ grid for different orderings and stencil sizes, $g=1, 2, 3$, and $4$.}
  \label{fig:xeon_gol3d_n64}
\end{figure}

\begin{figure}[ht]
  \centering
  \includegraphics[width=\linewidth]{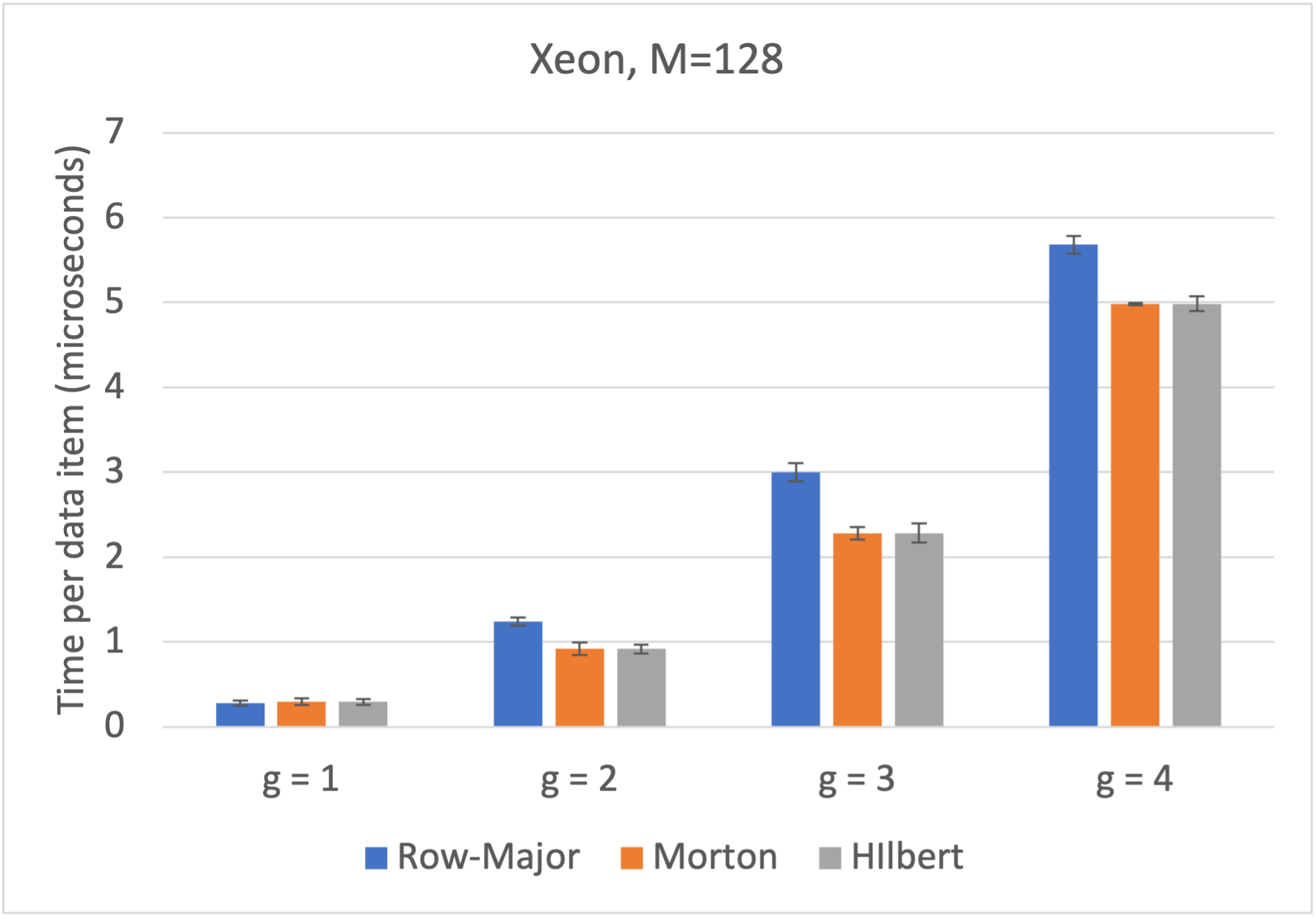}
  \caption{Time per data item for 10 iterations for a $128\times 128\times 128$ grid for different orderings and stencil sizes, $g=1, 2, 3$, and $4$.}
  \label{fig:xeon_gol3d_n128}
\end{figure}

\begin{figure}[ht]
  \centering
  \includegraphics[width=\linewidth]{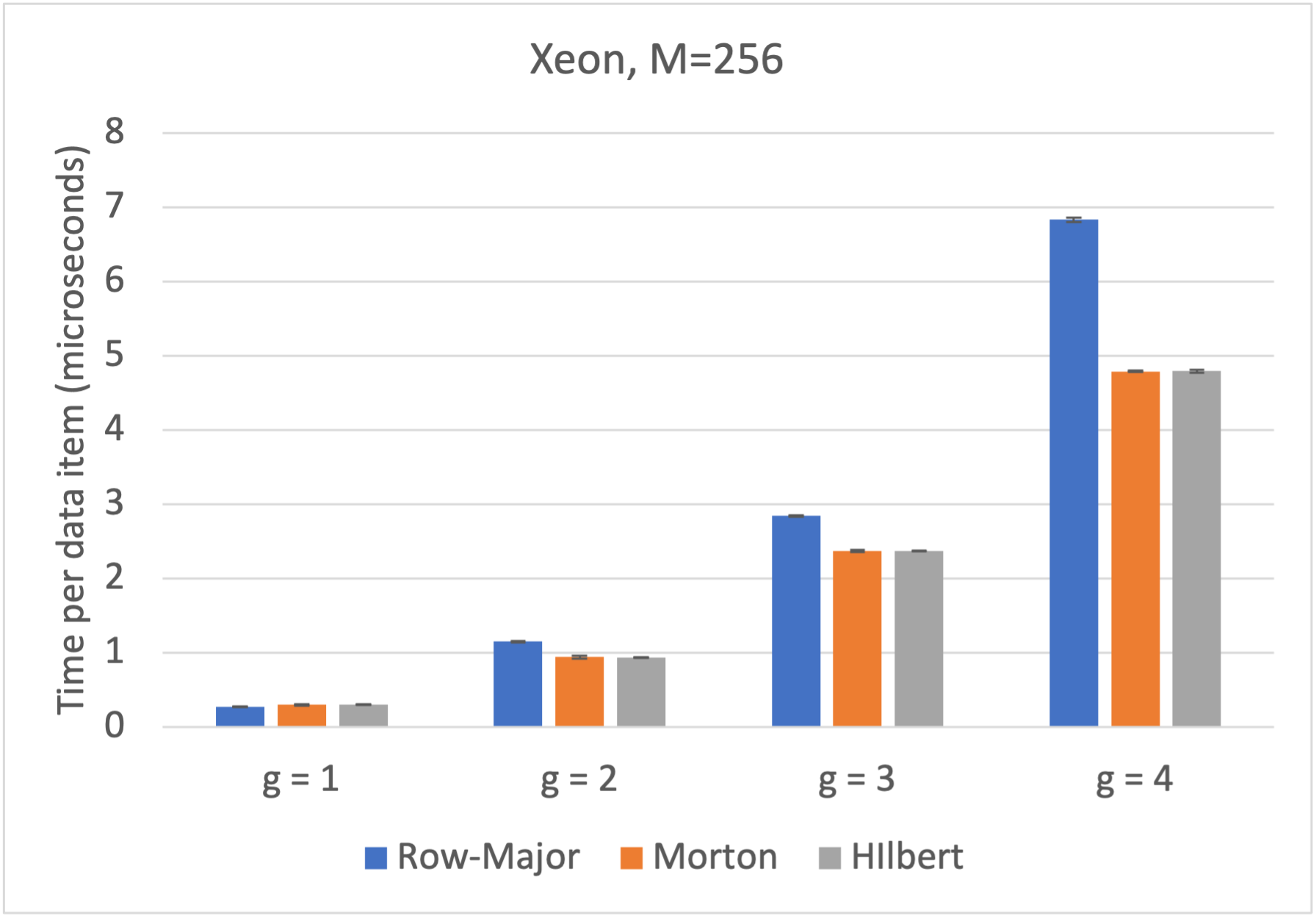}
  \caption{Time per data item for 10 iterations for a $256\times 256\times 256$ grid for different orderings and stencil sizes, $g=1, 2, 3$, and $4$.}
  \label{fig:xeon_gol3d_n256}
\end{figure}

Figures~\ref{fig:xeon_gol3d_n64}-\ref{fig:xeon_gol3d_n256} show the timings for running {\it gol3d} on the Xeon processor.
Although for the Epyc processor the timings were the same for the three orderings, it can be seen that for the Xeon processor the Morton and Hilbert orderings are faster than row-major.

\begin{figure*}[ht]
  \includegraphics[width=0.4\linewidth]{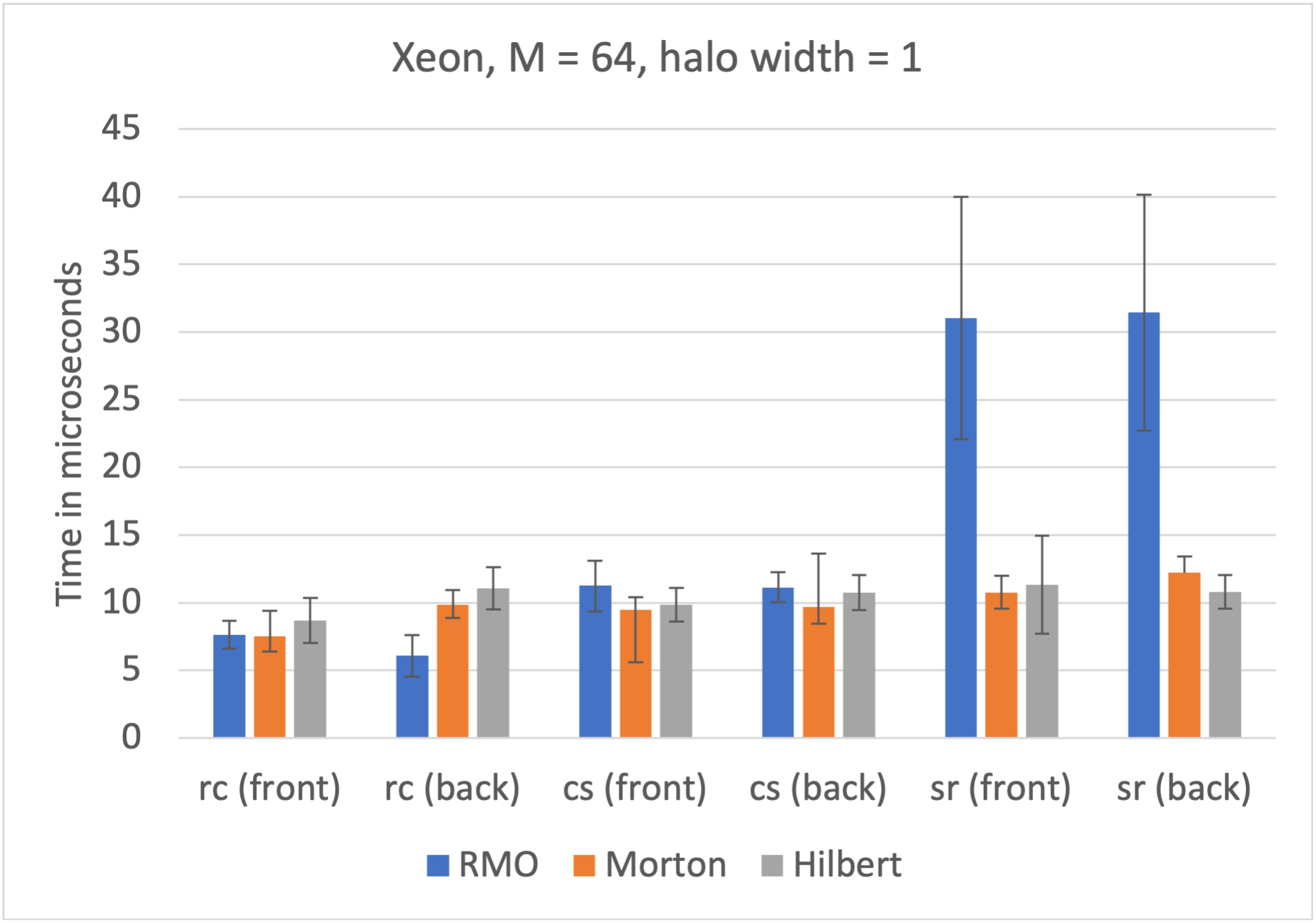}
  \qquad
  \includegraphics[width=0.4\linewidth]{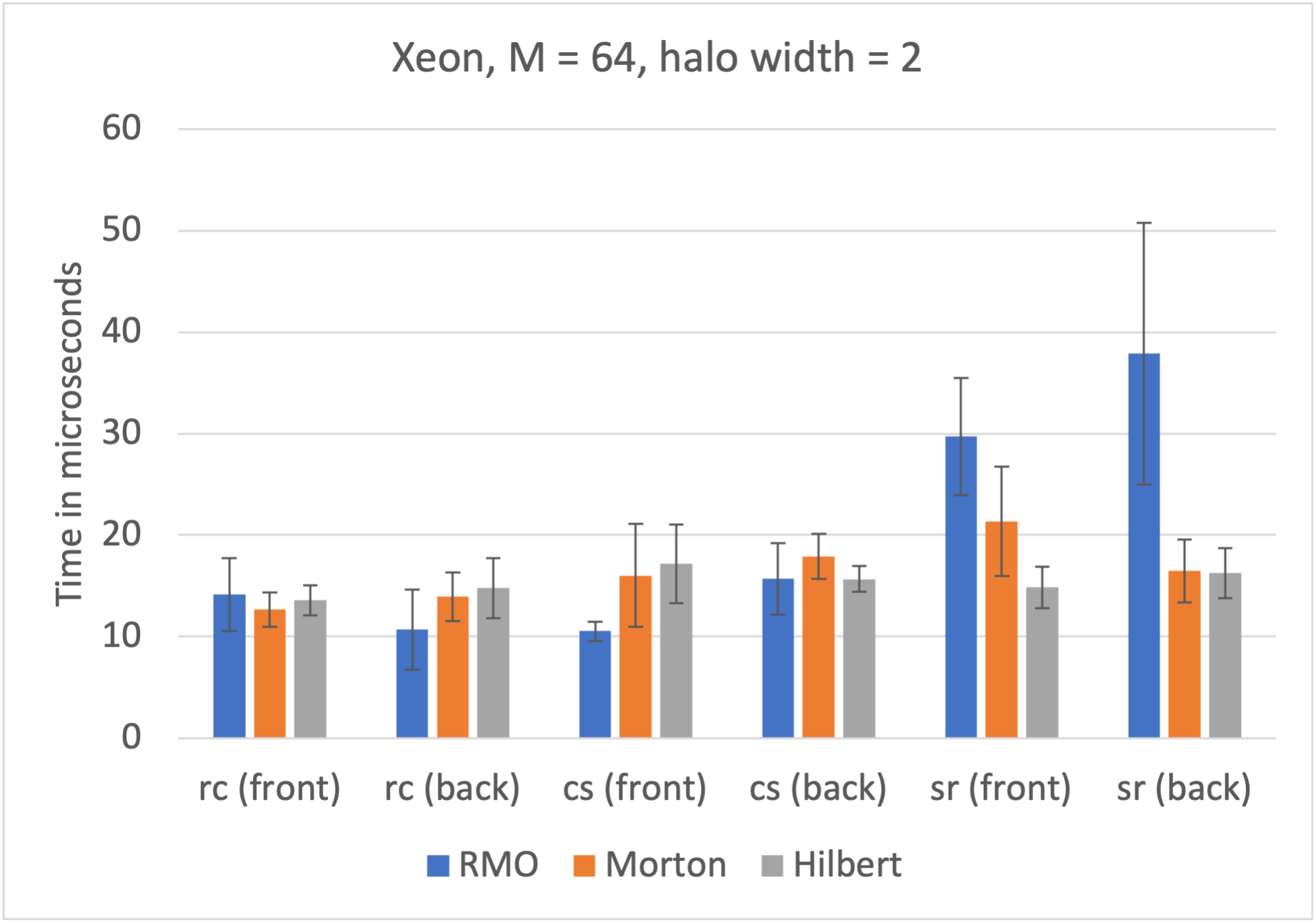}  \\
    \includegraphics[width=0.4\linewidth]{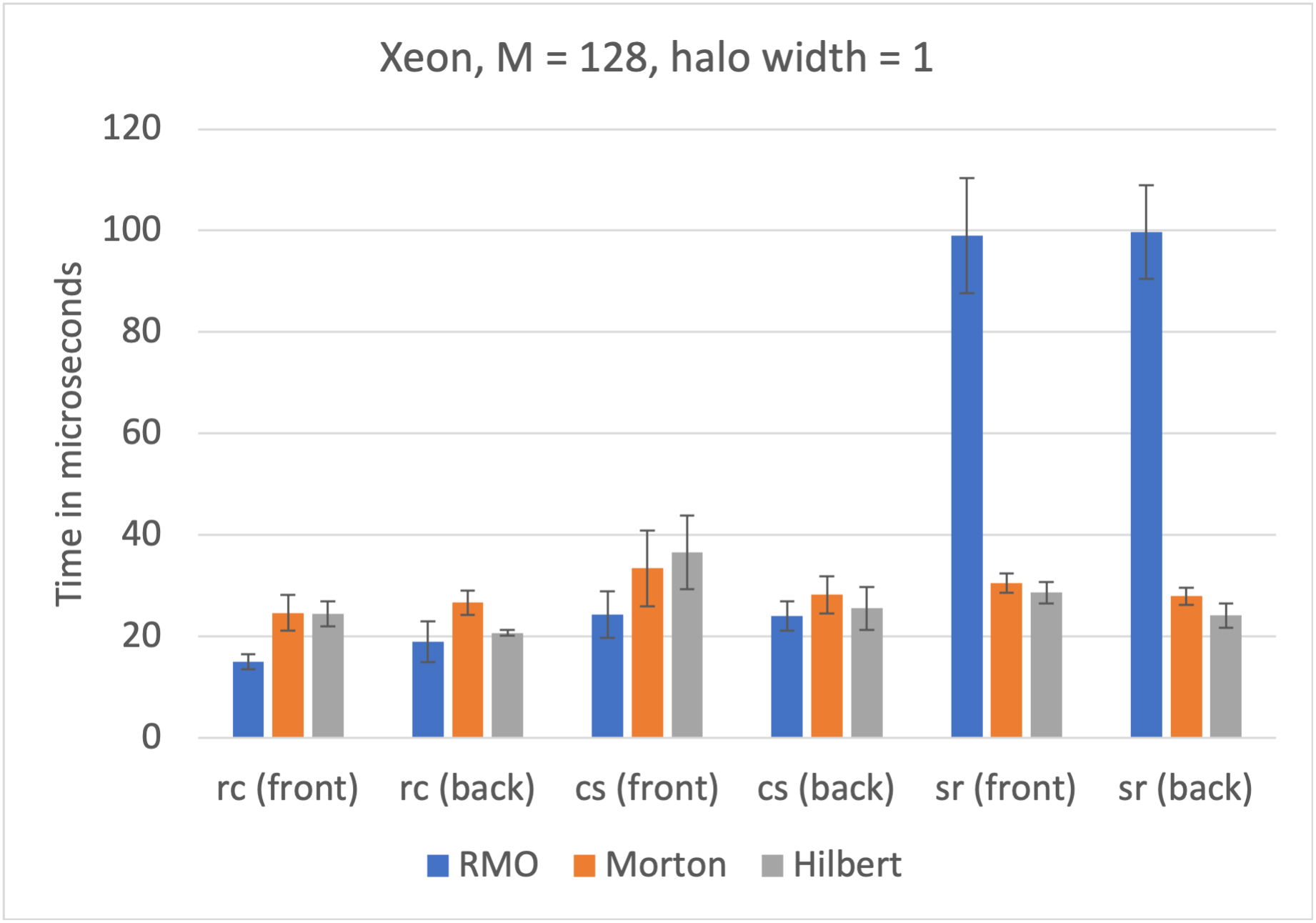}
  \qquad
  \includegraphics[width=0.4\linewidth]{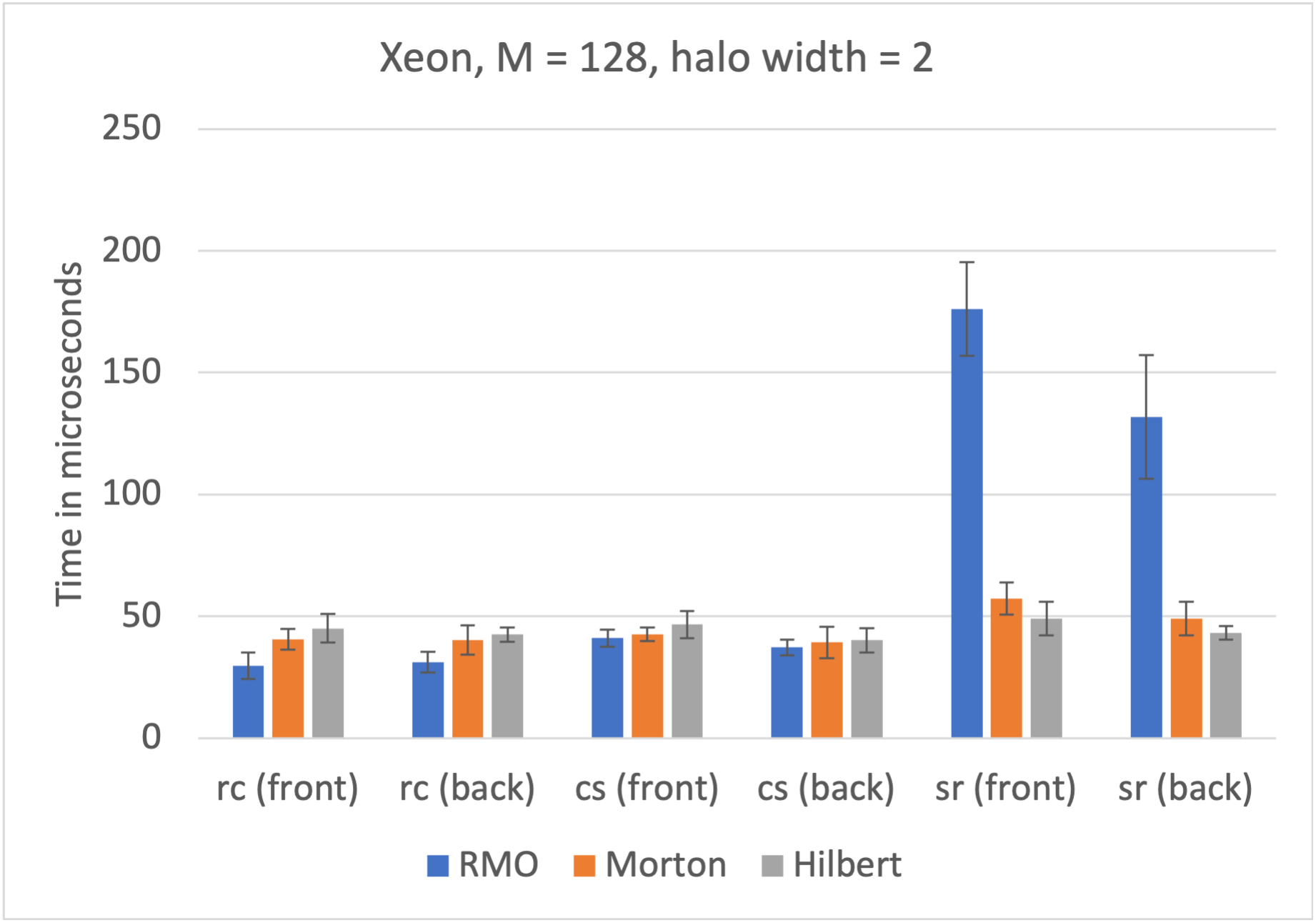} \\
      \includegraphics[width=0.4\linewidth]{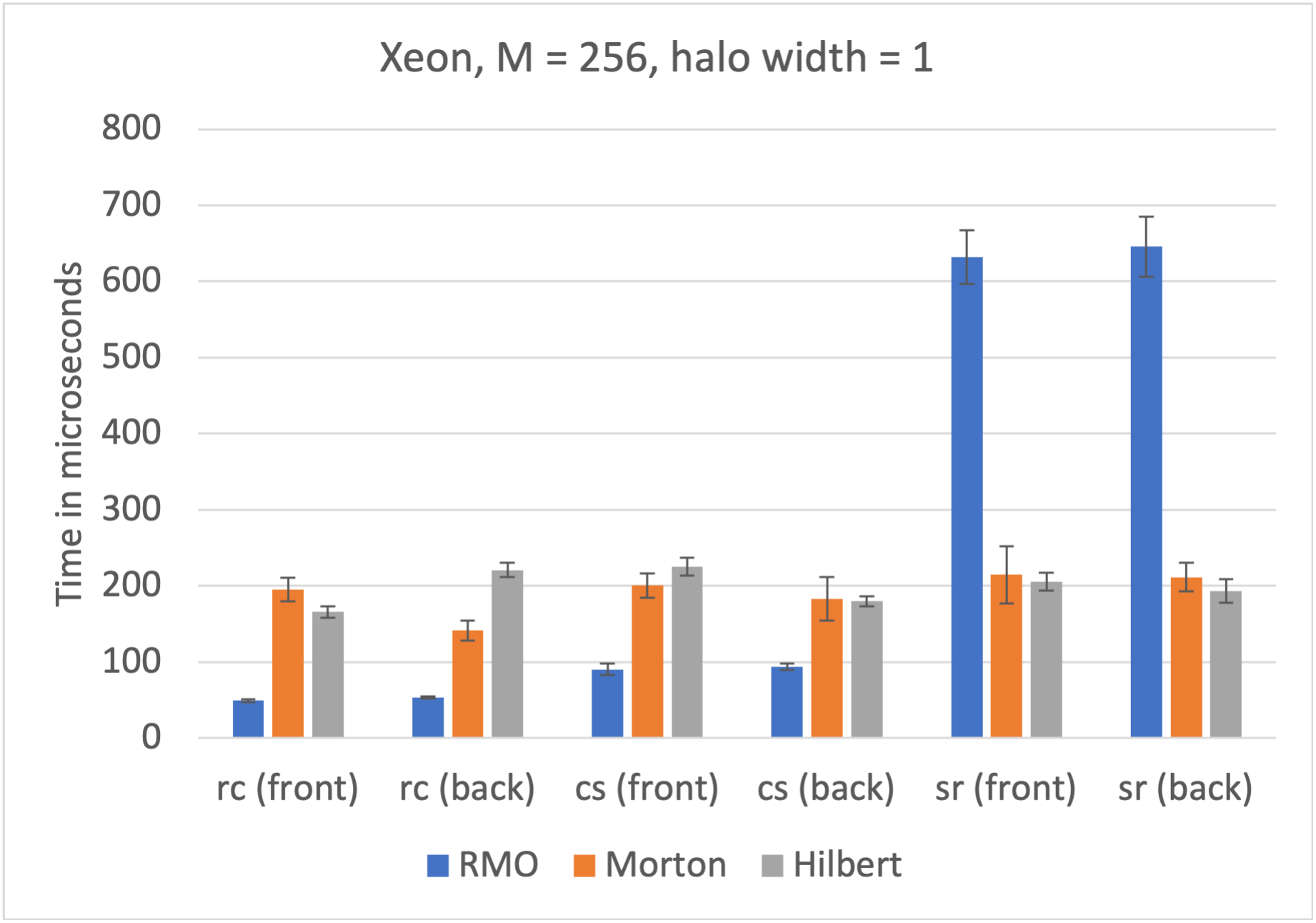}
  \qquad
  \includegraphics[width=0.4\linewidth]{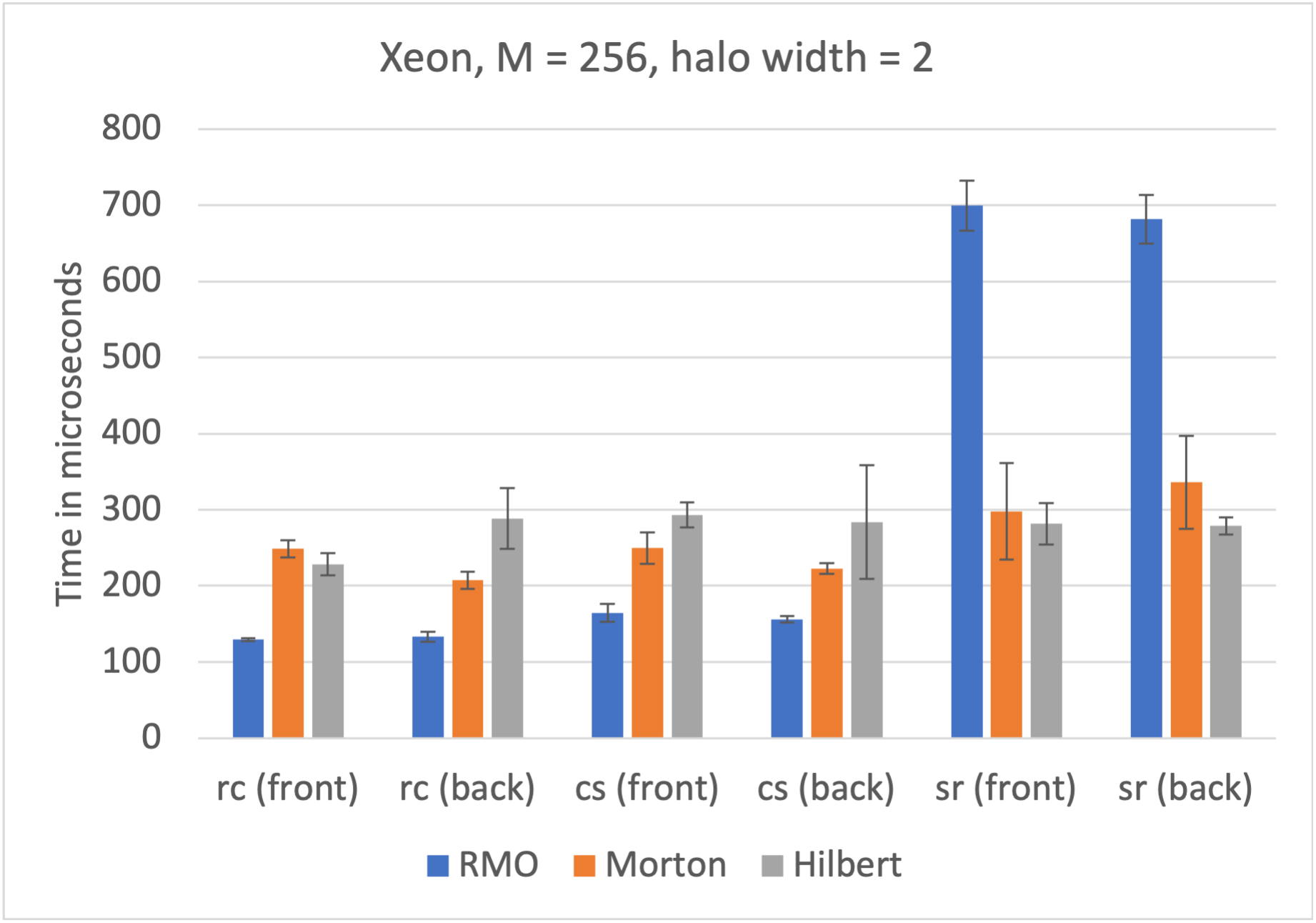}  \caption{Xeon results: time to buffer surfaces, for halo widths 1 and 2 and problem sizes 64, 128, and 256.}
  \label{fig:xeon_buffer}

\end{figure*}

Figure~\ref{fig:xeon_buffer} shows the times to write the surface data to the communication buffer. The results are broadly  the same as for the Epyc processor, with Morton and Hilbert orderings being faster to write to the communication buffer than the row-major case for the slab-row surfaces.

\section{Analysis and Interpretation}
\label{section:analysis}
The timings in Fig.~\ref{fig:epyc_buffer} broadly show that for the row-column (rc) and column-slab (cs) surfaces, a row-major ordering results in a faster time to write data into the corresponding communication buffer. However, for the slab-row (sr) surfaces a row-major ordering is slower than both Morton and Hilbert orderings, with the latter being fastest. In addition, for row-major ordering, buffering is fastest for rc surfaces and slowest for sr surfaces. The timing results for row-major ordering are fastest for rc surfaces and slowest for sr surfaces because moving along a row accesses data with stride 1, whereas moving with constant column or slab index accesses data with stride $n$ and $n^2$, respectively. The better performance for the Morton and Hilbert cases when buffering the sr surfaces is attributed to their better locality properties which vary only mildly as the different surfaces are buffered.

\begin{figure}[ht]
  \centering
  \includegraphics[width=0.8\linewidth]{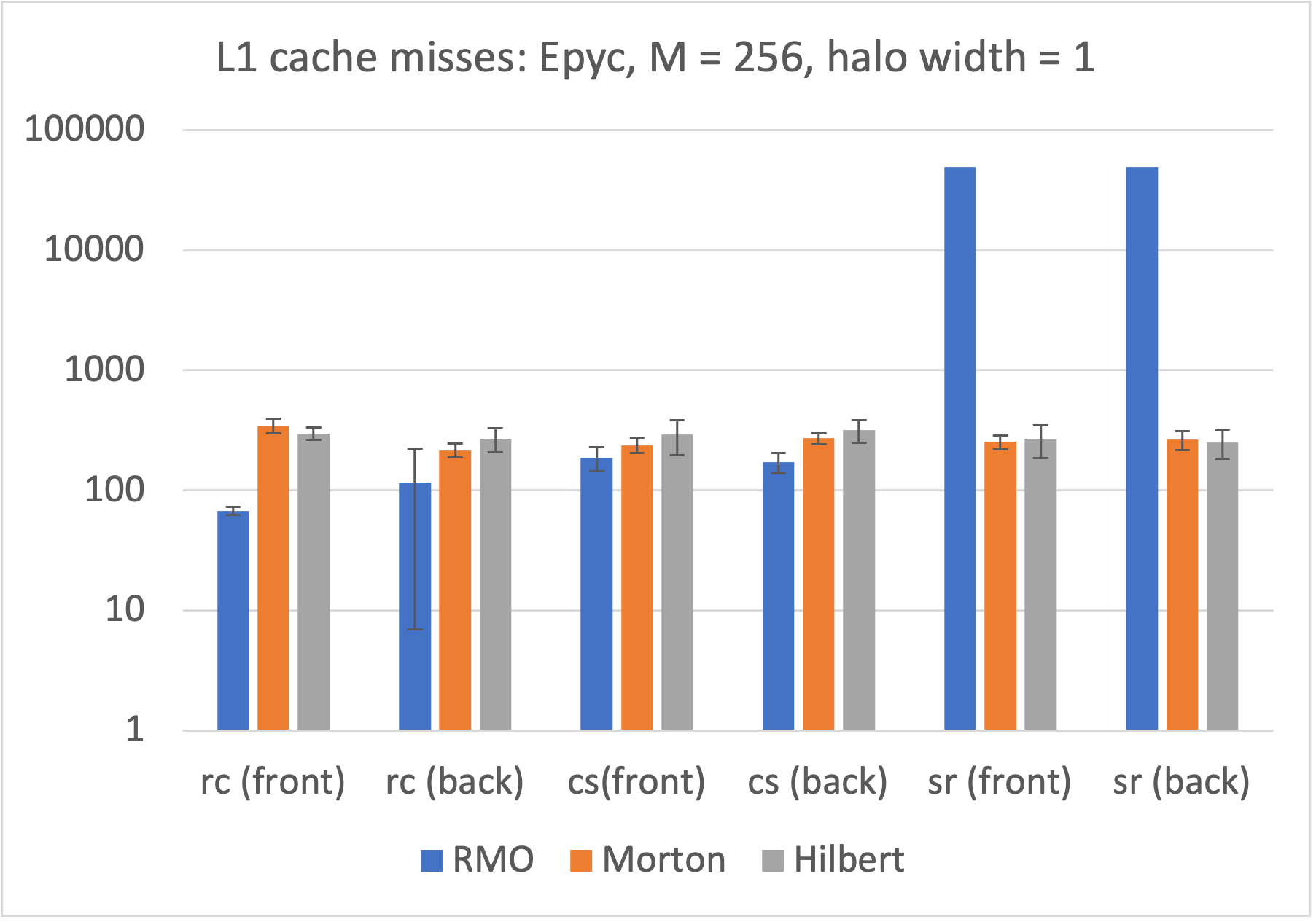}
  \caption{Epyc results: number of L1 data cache misses for a $256\times 256\times 256$ grid for halo size 1.}
  \label{fig:L1misses}
\end{figure}

\begin{figure}[ht]
  \centering
  \includegraphics[width=0.8\linewidth]{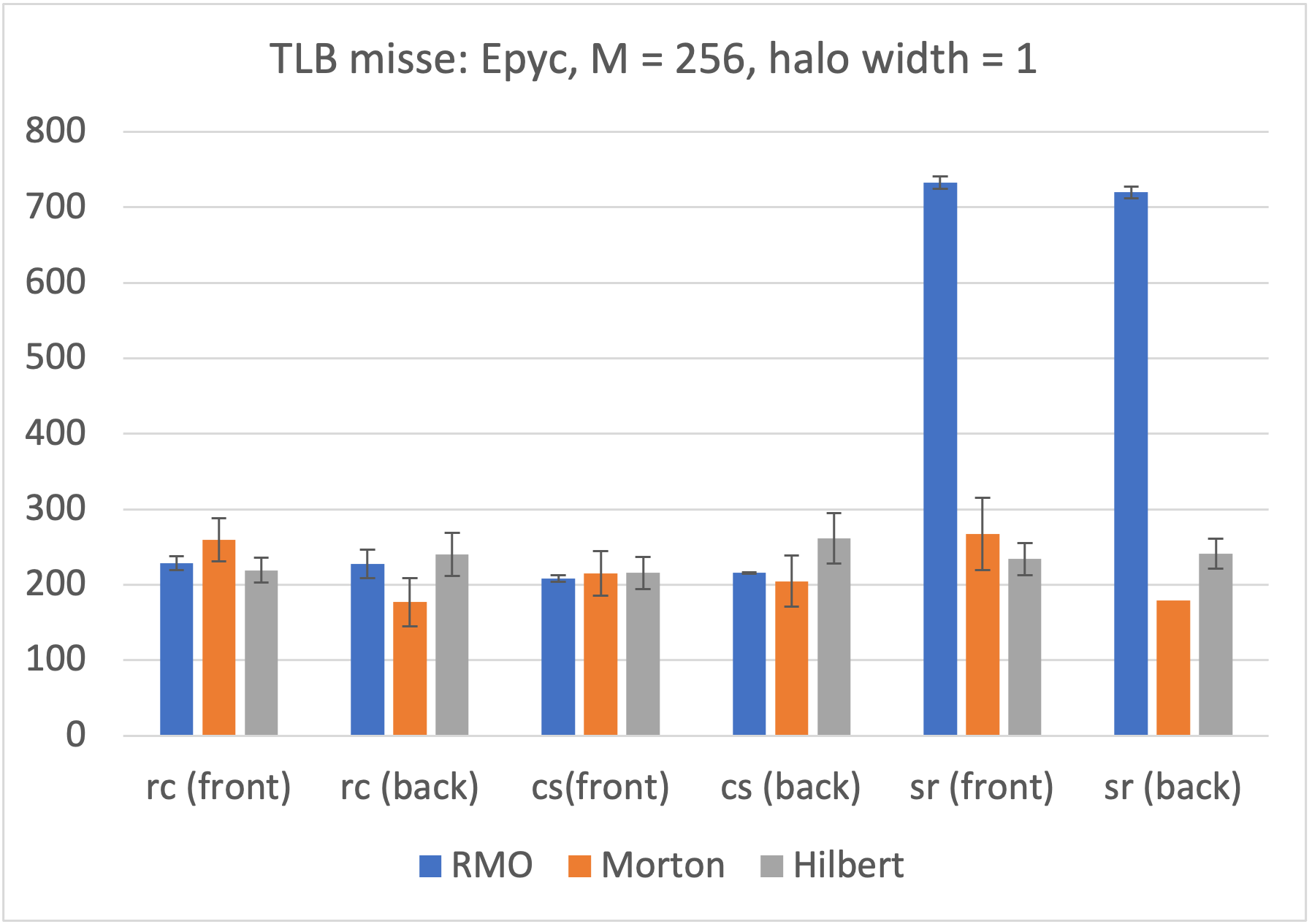}
  \caption{Epyc results: number of data TBL misses for a $256\times 256\times 256$ grid for halo size 1.}
  \label{fig:DTLBmisses}
\end{figure}

\begin{figure}[ht]
  \centering
  \includegraphics[width=0.8\linewidth]{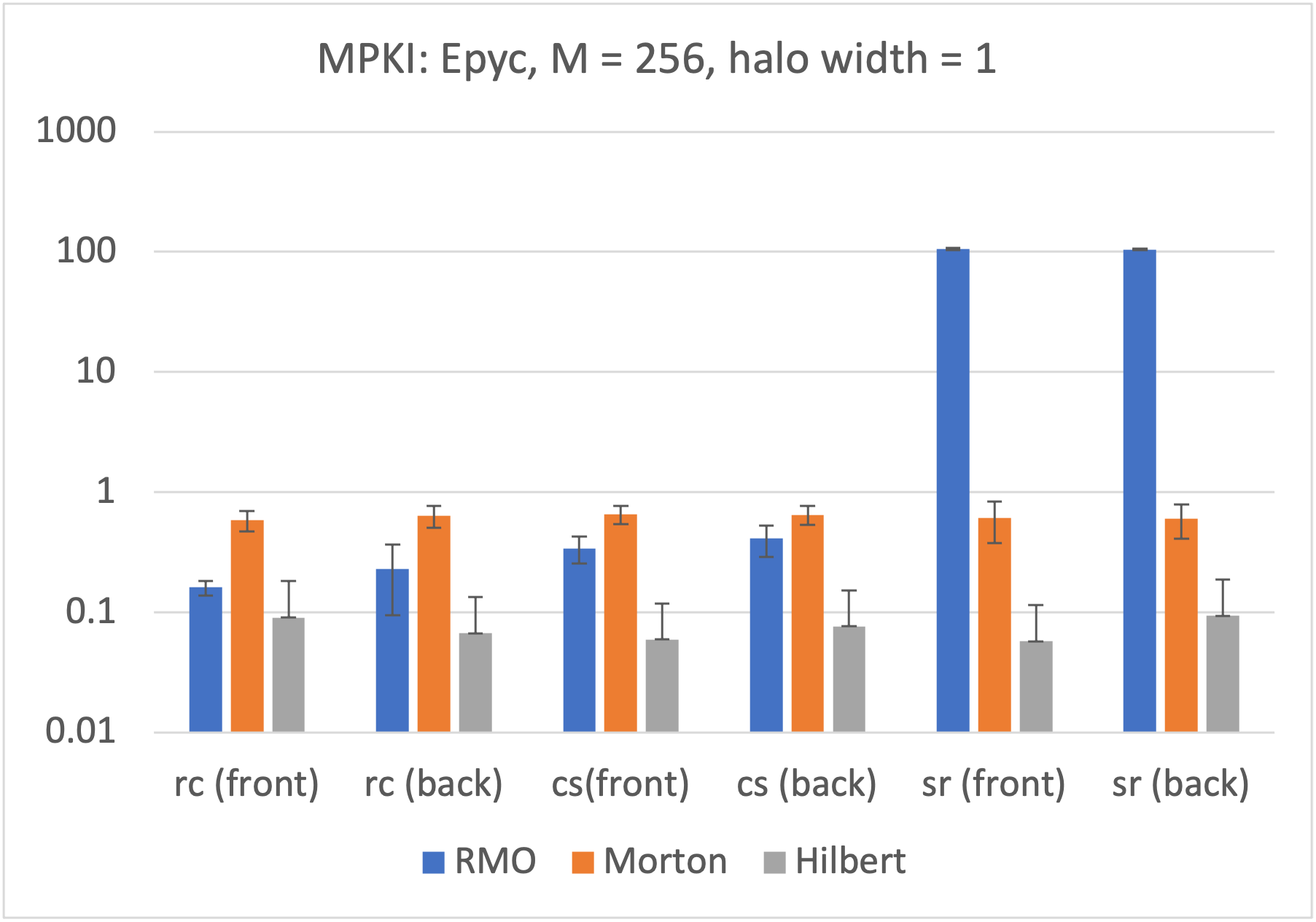}
  \caption{Epyc results: misses per kilo instruction for a $256\times 256\times 256$ grid for halo size 1.}
  \label{fig:MPKI}
\end{figure}

\begin{figure}[ht]
  \centering
  \includegraphics[width=0.8\linewidth]{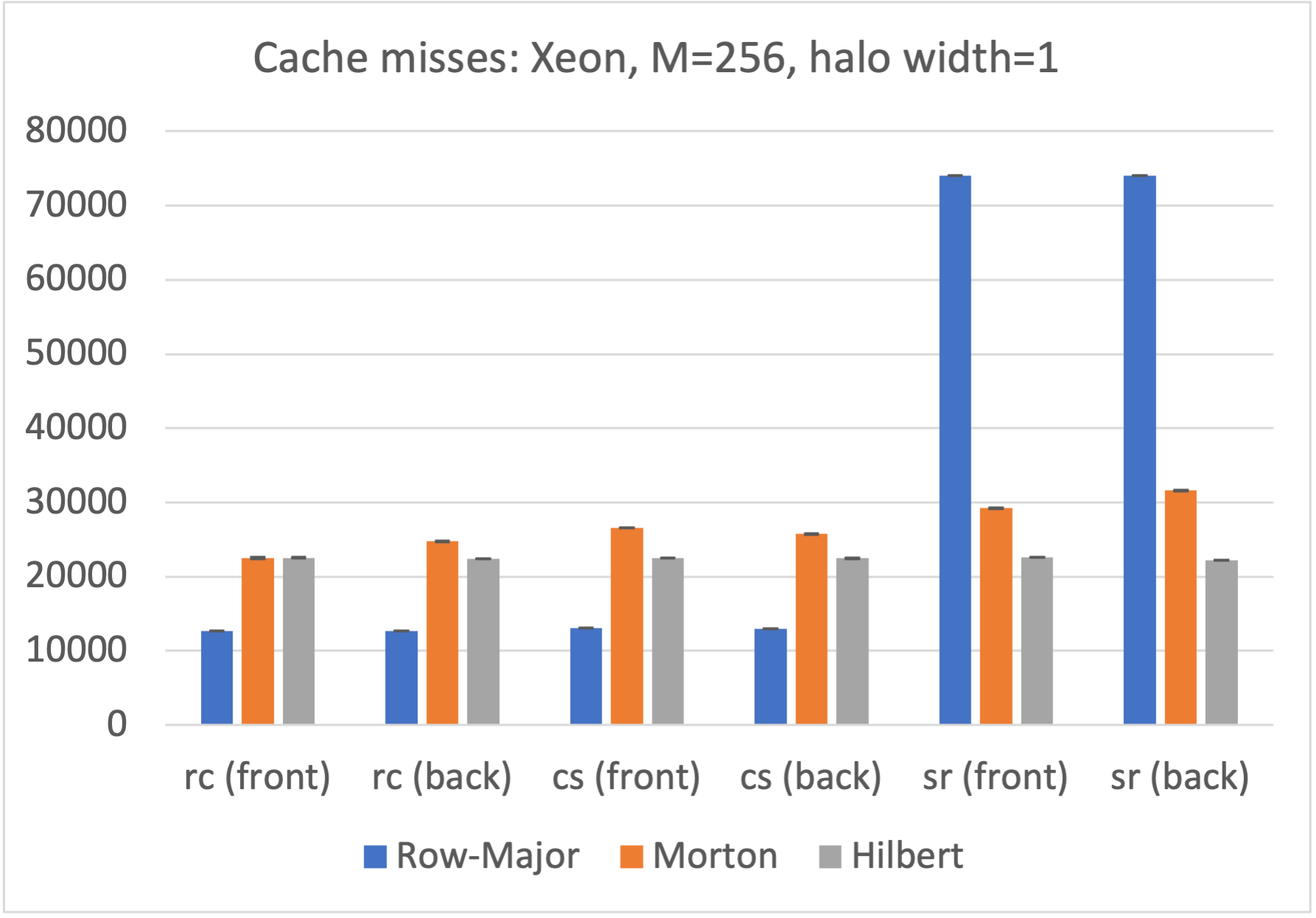}
  \caption{Xeon results: number of data cache misses for a $256\times 256\times 256$ grid for halo size 1.}
  \label{fig:Xeon_L1misses}
\end{figure}

The impact of the locality properties of an ordering when writing a surface to a communication buffer is illustrated in Figs.~\ref{fig:L1misses} -- \ref{fig:MPKI}, which show the number of L1 data cache misses, the data translation lookaside buffer (TLB) misses, and the misses per kilo instruction (MPKI) for $M=256$ and halo width 1, corresponding to the timing results in the lower left plot in Fig.~\ref{fig:epyc_buffer}. Note that the y-axis in Fig.~\ref{fig:L1misses} is logarithmic in order to show the large number of cache misses for the sr surfaces: $49601\pm 1310$ and $49136\pm 1766$ for the front and back surfaces respectively. Figure~\ref{fig:MPKI} shows that the MPKI metric is smallest for the Hilbert ordering ($<0.1$) and higher for the Morton ordering ($\approx 0.6$). For the row-major ordering, MPKI is approximately 0.2 and 0.38 for rc and cs surfaces, respectively, but rises to approximately 105 for the sr surfaces.
These data were generated using the perf\_event Linux kernel library, which is used to access hardware counters on the processor. The apparently large number of cache misses for the Xeon processor, shown in Fig.~\ref{fig:Xeon_L1misses}, may be due to an anomalous interaction between the perf\_event library and Xeon's proprietary cache architecture. However, the relative numbers of cache misses are similar to the results for Epyc, shown in Fig.~\ref{fig:L1misses}. 

\begin{figure}[ht]
  \centering
  \includegraphics[width=0.8\linewidth]{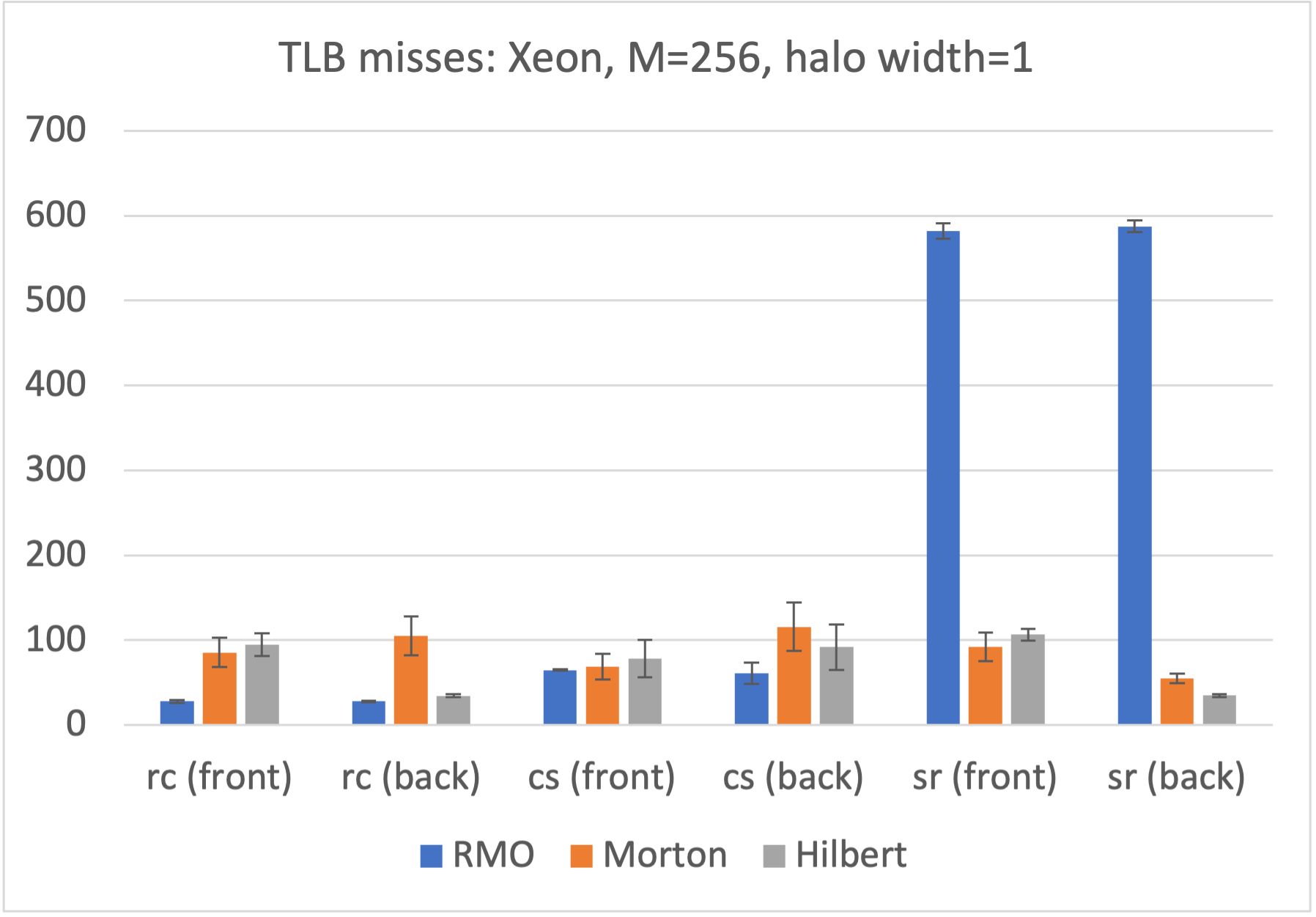}
  \caption{Xeon results: number of data TBL misses for a $256\times 256\times 256$ grid for halo size 1.}
  \label{fig:Xeon_DTLBmisses}
\end{figure}


\section{Conclusions and Future Work}
\label{section:conclusions}
First, we provide conclusions, then discuss potential future work.
\subsection{Conclusions}
This research has investigated how the locality properties of different data orderings affect the runtime of 3D stencil-based computations and the time to copy surface data into communication buffers. Results for AMD Epyc and Intel Xeon processors have been presented. The experimental timing results have been interpreted in terms of the reported number of cache and TLB misses. 

Figures~\ref{fig:xeon_gol3d_n64}--\ref{fig:xeon_gol3d_n256} show that the volume computations performed on a Xeon Gold processor are faster when performed in the space-filling data structures as opposed to a traditional row-major ordering of a data cube.  This is not so apparent for the AMD Epyc processor (see Figures~\ref{fig:epyc_gol3d_n64}--\ref{fig:epyc_gol3d_n256}). Further study of the differences in cache behaviors is warranted to better elucidate this interesting result.  

Previous and ongoing work has highlighted the scope for optimising the packing and unpacking of communication buffers in MPI~\cite{TEMPI,yaksa}. The cost of packing and unpacking surface data, which is a prerequisite to using MPI for neighbor communication, shows the advantages of using space-filling curves when packing/unpacking communication buffers.  The results in Figures~\ref{fig:epyc_buffer} and \ref{fig:xeon_buffer} show that, for both the Xeon and Epyc processors, there is a substantial advantage with space-filling curves for the slab-row surfaces, with nominal disadvantages for the other surfaces for certain parametrizations explored.  Overall, there is a significant net benefit from the space-filling data structure approach when both computation and communication is considered.

\subsection{Future Work}
\label{sec:future_work}
There are a number of directions for further inquiry that appear promising, given results thus far.  First, these considerations apply to the use of the space-filling computations themselves:
\begin{itemize}
\item Different processors will manage their caches in sufficiently different ways that exploring other processors and variations of the AMD and Intel processors appears valuable.
\item Further optimization of the 3D transformations between row-major and space-filling orderings should be investigated.  Prior work has shown that efficient incremental indexing is possible in 2D.  At present, the 3D transformations are all precomputed, implying costs for memory bandwidth and cache utilization.  The presence of a lookup table could have differential impacts on performance on different processors with varied cache policies and architectures.
\item This work should be explored with GPU-accelerated data cubes for the purpose of understanding how the space-filling data structures work well in such scenarios.  Evidently, greater internal bandwidth in GPUs may lessen the value of space-filling curves.  It need not reduce the value of managing the data transfers between GPUs, as noted below. 
\item Use of space-filling structures on more complex subdomains is of interest, beyond data cubes.  
\item Exploration of C++, DSL-based, or other abstractions to simplify the efficient use of space-filling data structures without losing performance should be of value.
\item Potential for introducing hardware instructions  that reduce the cost of transforming to/from space-filling addressing in both CPU and GPUs is of potential value (e.g., ISA extensions for x86-64 or ARM, including vectorized and incremental indexing shortcuts \cite{Valsalam98}). 
\item Coping with non-powers-of-2 efficiently, which has been done in previous efforts with space-filling curves, needs to be done here as well to support general problem sizes \cite{Valsalam2002}.
\end{itemize}
Interaction with MPI communication suggests these additional studies:
\begin{itemize}
\item {\sloppy Revisit the communication performance in the GPU-acceler\-ated scenarios in which the data cube resides in  GPU memory.  Both current and near-future coherent memory cases should be considered.  Packing/unpacking of buffers for halo codes in GPU-accelerated systems comprise an important overhead presently, and reducing this overhead through a better overall data structure would be valuable to improving MPI+X performance.}
\item Explore if pre-compiled derived MPI datatypes can work as effectively as the hand-packed performance shown in this paper (runtime datatypes are likely to be much slower than hand packing and unpacking).
\end{itemize}

\begin{acks}
This work was performed with partial support from the National Science
Foundation under Grants Nos.~ 
CCF-1822191, CCF-1821431, OAC-1923980, OAC-1549812, OAC-1925603, and OAC-2201497 and the U.S. Department of Energy's National Nuclear Security Administration (NNSA) under the Predictive Science Academic Alliance Program (PSAAP-III), Award DE-NA0003966.

Any opinions, findings, and conclusions or recommendations expressed in this material are those of the authors and do not necessarily reflect the views of the 
National Science Foundation and the U.S. Department of Energy's National Nuclear Security Administration.

The authors acknowledge use of research infrastructure of the UTC SimCenter in preparing this research.
\end{acks}

\bibliographystyle{ACM-Reference-Format}
\bibliography{CUPECS-2023-19}

\end{document}